\documentclass{aa}
\usepackage{graphicx}

\newcommand{\msun}{\mbox{$M_{\odot}$}}

\newcommand{\bv}{\mbox{$B\!-\!V$}}
\newcommand{\vi}{\mbox{$V\!-\!I$}}
\newcommand{\vk}{\mbox{$V\!-\!K$}}
\newcommand{\gz}{\mbox{$g\!-\!z$}}
\newcommand{\feh}{\mbox{$[$Fe/H$]$}}

\begin{document}
\title{Globular Clusters in NGC~4365: New $K$-band Imaging and a
 Reassessment of the Case for Intermediate-age Clusters
  \thanks{Based on observations collected at the European Southern
          Observatory, Chile under programmes 70.B-0289A and 71.B-0303A, 
	  and with the NASA/ESA
    \emph{Hubble Space Telescope}, obtained at the Space Telescope
	  Science Institute, which is operated by the Association
	  of Universities for Research in Astronomy, Inc., under
	  NASA contract NAS 5-26555}
   \fnmsep
  \thanks{Table 3 is only available in electronic form at the CDS via 
      anonymous ftp to cdsarc.u-strasbg.fr (130.79.128.5) or via 
      http://cdsweb.u-strasbg.fr/cgi-bin/qcat?J/A+A/}
}

\author{S. S. Larsen
        \inst{1} 
	\and 
	J. P. Brodie 
	\inst{2}
	\and 
	J. Strader
	\inst{2}
}

\institute{European Southern Observatory, ST-ECF,
  Karl-Schwarzschild-Str. 2, D-85748 Garching b.\ M{\"u}nchen, Germany
  \and
  UCO / Lick Observatory, 1156 High Street,
  University of California, Santa Cruz, CA 95064, USA
}

\offprints{S.\ S.\ Larsen, \email{slarsen@eso.org}}

\date{Received 09.05.2005; Accepted 03.08.2005}

\abstract{
  We study the globular cluster (GC) system of the Virgo giant elliptical
galaxy \object{NGC~4365}, using new wide-field $K$-band imaging from the 
ESO 3.5 m 
New Technology Telescope, archive $V$ and $I$ imaging from FORS1 on the ESO 
VLT and HST/WFPC2+ACS data.  As in most other large ellipticals, the 
GC colour distribution has (at least) two peaks, but the colours of the red 
GCs appear more strongly weighted towards intermediate colours compared to 
most other large ellipticals and the integrated galaxy light.  The 
intermediate-color/red 
peak may itself be composed of two sub-populations, with clusters of
intermediate colours more concentrated towards the centre of the galaxy than 
both the blue and red GCs.  Nearly all GCs in our sample fall along 
a well-defined narrow sequence in the (\vk,\vi) two-colour diagram, with
an offset towards red \vk\ and/or blue \vi\ colours compared to simple stellar 
population models for old ages. This has 
in the past been interpreted as evidence for an intermediate-age population of 
GCs.  The offset is however seen for nearly all metal-rich clusters within 
the $5\arcmin\times5\arcmin$ SOFI field, not just those of intermediate colours.
We combine our $VIK$ data with previously published spectroscopy
resulting in a sample of 25 GCs with both spectroscopy and 
photometry.  The differences between observed and model colour-metallicity 
relations are consistent with the offsets observed in the two-colour diagram, 
with the metal-rich GCs being too red (by $\approx0.2$ mag) in \vk\ and too 
blue (by $\approx0.05$ mag) in \vi\ compared to the models at a given
metallicity.  These offsets cannot easily be explained as an effect of 
younger ages.  We further compare the 
colour-metallicity relation for GCs in NGC~4365 with previously published data 
for \object{NGC~3115} and the Sombrero galaxy, both of which are believed from 
spectroscopic studies to host exclusively old GC populations, and find
the colour-metallicity relations for all three galaxies to be very similar.
We review the available evidence for
intermediate-age GCs in NGC~4365 and conclude that, while this cannot be 
definitively ruled out, an alternative scenario is more likely whereby
all the GCs are old but the relative number of 
intermediate-metallicity GCs is greater than typical for giant ellipticals.
The main obstacle to reaching a definitive conclusion is the lack 
of robust calibrations of integrated spectral and photometric 
properties for stellar populations with near-solar metallicity. 
In any case, it is puzzling that the significant intermediate-colour GC
population in NGC~4365 is not accompanied by a corresponding shift of 
the integrated galaxy light towards bluer colours.

}

\titlerunning{Globular Clusters in NGC~4365}

\maketitle

\section{Introduction}

  One of the main motivations for the numerous studies of extragalactic
globular cluster (GCs) carried out over the past 1-2 decades 
is the expectation that GCs can be used as tracers of (major) star formation 
episodes in their host galaxies. To first order, it is probably safe to 
assume that the apparent ubiquity of bi-modal colour distributions in GC 
systems (Larsen et al.\ \cite{lar01} - L2001; Kundu \& Whitmore \cite{kw01}; 
Gebhardt \& Kissler-Patig \cite{gk99}) reflects
corresponding multiple episodes/mechanisms of star formation. 
Assuming a one-to-one relation between ``field'' star populations and GCs 
in general may be dangerous, however. Metal-poor and 
metal-rich GCs appear to form (or survive!) with different efficiencies 
relative to field 
stars of the corresponding metallicities, as shown directly from photometry 
of resolved stars in the nearby giant elliptical \object{NGC~5128} (Cen A) by
Harris et al.\ (\cite{hhp99}). 
  Harris \& Harris (\cite{hh02}) find that the specific frequency of blue
GCs in NGC~5128 is 3--5 times higher than for the red ones.
In the giant Fornax elliptical \object{NGC~1399},
Forte et al.\ (\cite{ffg05}) have shown that the integrated galaxy
colours and GC colour distributions suggest specific frequencies of
$3.3\pm0.3$ for the red GCs and $14.3\pm2.5$ for the blue ones with
respect to field stars of the same metallicities, i.e.\ there are about
4 times as many metal-poor GCs per metal-poor field star. While this
assumes that the GCs and field stars follow the same radial
distributions, the results are consistent with those for NGC~5128. This makes 
the integrated colours of early-type galaxies generally redder than the 
average colours of their GC systems.  It is 
also worth keeping in mind that while some 2/3 of the known GCs in the 
Milky Way appear to be associated with the stellar halo (Zinn \cite{zinn85};
Harris \cite{har96}), that component 
accounts for only $\sim1$\% of the stellar mass in our Galaxy. The disk
or our galaxy is about an order of magnitude more massive than the 
bulge (Dehnen \& Binney \cite{db98}), so naively scaling from the $\sim50$ 
bulge GCs one might expect 
about 500 GCs to have formed in the disk over its lifetime, or about
one every 20 Myr on average.  Few if any 
young clusters with masses greater than $10^5$ $\msun$ have in fact been
identified so far in our Galaxy (the cluster Westerlund 1 may be one
example, see Clark et al.\ \cite{cl05}), although they are frequently
observed in many starbursts, mergers, and even some apparently normal 
spirals (see Larsen \cite{lar05} for a review).  Real physical differences 
in the cluster mass functions can be difficult to disentangle from 
size-of-sample effects (Billett et al.\ \cite{bhe02}; Larsen \cite{larsen02}; 
Hunter et al.\ \cite{hun03}; Whitmore \cite{whit03}; 
Weidner et al.\ \cite{wkl04}; Gieles et al.\ \cite{gie05}) and it remains 
unclear 
whether or not formation of massive star clusters is favoured relative to 
low-mass ``open'' clusters under certain conditions (e.g.\ intense starbursts).
If we wish to use GCs as tracers of galaxy evolution, a 
necessary prerequisite will be to achieve a better understanding of the 
relationship between (massive) cluster formation/survival and star 
formation in general.

  Among the GC systems of early-type galaxies, a particularly puzzling case 
is NGC~4365.  The galaxy itself appears to be a typical giant elliptical 
galaxy in the Virgo cluster. It is among the 10 brightest early-type galaxies 
in Virgo (Binggeli et al.\ \cite{bin85}), but 
appears otherwise unremarkable.  Tonry et al.\ (\cite{ton01}) derived a 
distance modulus of $m-M = 31.55\pm0.17$ from surface brightness fluctuation 
(SBF) measurements, placing NGC~4365 about 0.5 mag ($\sim4$ Mpc) behind
the \object{M49} subcluster in the direction 
of the Virgo W cloud.  NGC~4365 has a kinematically decoupled 
core (KDC; Surma \& Bender \cite{sb95}), but this is not unusual among 
early-type galaxies and no relation has been found between the presence of 
KDCs and the overall properties of GC systems (Forbes et al.\ \cite{for96b}). 
Surma \& Bender derived a mean age of $7\pm1$ Gyr for the 
stellar population in NGC~4365 but this has later been revised to about 
$\sim14$ Gyr by Davies et al.\ (\cite{dav01}) who also showed (based on 
SAURON data) that there are no age gradients across the central region where 
the KDC is located.  The difference between the age estimates was mostly due 
to the use of different population synthesis models and 
in fact the H$\beta$ line-index measurements of Davies et al.\ were in 
good agreement with those of Surma \& Bender. Brodie et al.\ (\cite{bro05})
quote similar H$\beta$ line-index measurements for NGC~4365 to those
of Davies et al.\ and Surma \& Bender, but also include the H$\gamma$ and
H$\delta$ lines. Depending on which lines are used, these data suggest 
ages between 7 and 14 Gyrs compared with simple stellar models by Thomas et 
al.\ (\cite{tmk04}). A critical difficulty in the interpretation of these
results is the poor empirical constraints on simple stellar population
(SSP) models at high metallicities.
This problems also applies to the globular cluster system.

Only the innermost region ($0\farcs2\approx15$ pc) of NGC~4365 shows some 
evidence for a younger (by 3--4 Gyr) stellar population, but this accounts 
for no more than $\sim2.5\times10^6 L_{\odot}$ (Carollo et al.\ \cite{car97}).
In other words, the \emph{stellar} population in NGC~4365 appears to be 
uniformly very old, and if the KDC is the result of a merger event then this 
must have happened very long ago. Surma \& Bender (\cite{sb95}) argue
that a dissipationless (``dry'') merger of two early-type 
galaxies cannot have produced the observed kinematical properties of 
NGC~4365, because an infalling stellar system cannot avoid phase mixing
and thus would not be observable as a kinematically distinct component.

  In the context of these results, the GC system of NGC~4365 displays some 
quite remarkable characteristics. A rather large amount of data have now 
been collected, sometimes yielding contradictory results, so in the 
following we give a reasonably comprehensive review of the current status:

Ajhar et al.\ (\cite{aj94}) 
first noted that the colour distribution of GCs in NGC~4365 was relatively 
narrow compared to \object{NGC~4472}. This result has been confirmed by 
many studies based on imaging with the \emph{Hubble Space Telescope}
(Forbes \cite{for96}; Gebhard \& Kissler-Patig \cite{gk99}; Larsen et al.\
\cite{lar01}) and must now be regarded as fairly secure. Initially, no
evidence was found for bimodality in the colour distribution of NGC~4365 GCs, 
which appeared better fit by a single broad distribution with a centroid at 
intermediate colours. However, the
WFPC2 \vi\ photometry used in much of the early work on NGC~4365 was
only moderately sensitive to metallicity differences, and 
the colour distribution could still be composed of a blue peak at the
``normal'' location ($\vi\approx0.9$ or $\feh\approx-1.5$ for an old
stellar population) and a red peak shifted somewhat towards the blue with
respect to the $\vi\approx1.2$ peak observed in most other large ellipticals
(e.g. L2001).  Since optical colours are degenerate in age and metallicity, 
an unusually blue \vi\ colour for the red ``peak'' might be interpreted either 
as a result of lower metallicity, or younger age.  We show below 
(\S\ref{sec:peccol}) that 
bimodality is in fact visible even in the \vi\ colour distribution.  

The age-metallicity degeneracy can in principle be lifted by including
multiple broad-band colours, each with different relative sensitivities to
age and metallicity. Puzia et al.\ (\cite{puz02}; P02) combined $K$-band 
imaging of NGC~4365, obtained with the ISAAC instrument on the ESO VLT, with 
WFPC2 $V$ and $I$ imaging, and found the distribution of the NGC~4365 GCs in
a \vi\ vs.\ \vk\ two-colour diagram to be different from those in NGC~3115,
the Milky Way and M31, being shifted to bluer \vi\ and/or redder \vk\
colours.  From a comparison with various SSP
models, this was attributed to about 40\%-80\% of the clusters in their 
sample belonging to a population with ages in the 
range 2--8 Gyrs. However, spatial and photometric completeness did not permit 
strong constraints on the total number of candidate intermediate-age clusters.
Supplementing 
the P02 $VIK$ data with deep $U$-band imaging, Hempel \& 
Kissler-Patig (\cite{hem04b}) again found suggestions of an intermediate-age 
GC population in NGC~4365, but with even weaker constraints on the total 
number of clusters belonging to this population.

  In Larsen et al.\ (\cite{lar03}; hereafter Paper I) we presented Keck/LRIS
spectroscopy for 14 confirmed GCs in NGC~4365, including 10 
objects from P02. A comparison of Balmer line indices 
(H$\beta$, H$\gamma$ and H$\delta$) with SSP models by Thomas, 
Maraston \& Bender (\cite{tmb03}) and R.\ Schiavon seemed to confirm that 
some of the GCs had intermediate ages ($\sim3$--5 Gyrs). Since most of
the GCs were selected as having $VIK$ colours suggestive of intermediate 
ages, this spectroscopic study also did not allow an estimate of the actual 
fraction.  However, it was noted that the presence of a large number 
of intermediate-age GCs in NGC~4365 would be puzzling considering the 
uniformly old, luminosity-weighted age of the underlying stellar light. 
In Paper I we estimated that at most 5\% of the total stellar mass 
at any given location within the galaxy could belong to a 5 Gyr population 
without noticeable effect on the H$\beta$ measurements of Davies et al.\ 
(\cite{dav01}). For a younger population, the limit is even lower.
 
  Additional spectra were presented by Brodie et al.\ (\cite{bro05}; 
Paper II).  The candidates were selected from VLT/FORS1 $V$ and $I$ 
imaging, whose larger field of view allowed 
selection of a sample of 22 GCs. The P02 $K$ band data
only covered a small fraction of the FORS1 field, and the spectroscopic 
sample of Brodie et al.\ was thus selected without regard to the \vk\ 
colours of the clusters, although an effort was made to make sure the full 
range of \vi\ colours was covered.  Still, the Paper II sample included 12
objects with $K$-band photometry from P02. In contrast to the 
Paper I study, $H\beta$ and H$\delta$ line index measurements for all 
GCs in this new sample indicated \emph{old} ($\sim12$ Gyr) ages 
(again using the Thomas et al.\ (\cite{tmb03}) SSP models), although
significant scatter was present around the 12-Gyr isochrone.
The H$\gamma_F$ line index measurements were
suggestive of somewhat younger ages for both GCs and the integrated
galaxy light.  
Curiously, three clusters 
in common between the Paper I and Paper II samples all shifted towards older 
ages in the Paper~II data.  We return to this issue below 
(\S\ref{sec:spec_ages}).

  In this paper we present new $K$-band imaging of GCs in NGC~4365,
obtained with the SOFI imager on the ESO New Technology Telescope (NTT). 
We combine the new $K$-band data with $V$ and $I$ imaging from the FORS1 
instrument on the ESO Very Large Telescope for a new analysis of the
\vi, \vk\ two-colour diagram, independent of the work by
Puzia et al.\ (\cite{puz02}).  While less deep than the ISAAC imaging, 
the combined SOFI$+$FORS1 data cover a larger field of view,
allowing us to better constrain whether an intermediate-age
population (if present) is restricted to the central parts of the
galaxy. Furthermore, the SOFI data are deep enough to comfortably include 
all clusters for which spectroscopy was obtained in Paper I and II
so that a comparison of photometric and spectroscopic age indicators, as
well as other properties, can be carried out.  In order to keep our
analysis as model-independent as possible, we compare the NGC~4365 data
with previously published spectroscopy and photometry for NGC~3115 
(P02; Kuntschner et al.\ \cite{kunt02}) and
the Sombrero galaxy (Larsen et al.\ \cite{lar02}).  We also include 
F850LP ($z$) and F475W ($g$) band photometry from the ACS Virgo Cluster 
Survey (C{\^o}t{\'e} et al.\ \cite{cot04}), and compare the
F850LP, F475W, F555W and F814W HST data for GCs in NGC~4365 with similar
data for three other large Virgo ellipticals, \object{NGC~4406}, 
\object{NGC~4649} and \object{NGC~4486}, which have about the same luminosity.

\section{Observations and data reduction}

\label{sec:data}

A realistic assessment of the limitations of the photometry will be
important for the discussion later on. In the following we therefore discuss 
our observational strategy and data reduction in some detail. 

\subsection{SOFI $K$-band imaging: data}

\begin{table}
\caption{\label{tab:obslog}Log of SOFI $K$-band observations}
\begin{tabular}{lccccc} \hline
  Date     & N(exp) &  Seeing    & \multicolumn{3}{c}{Airmass} \\ 
           &        &            &   Mean  &   Min  &  Max \\ \hline
2004-04-04 &   40   & $1\farcs4$ &  1.247  & 1.244  & 1.253 \\
2004-04-05 &  146   & $1\farcs0$ &  1.406  & 1.244  & 1.936 \\
2004-04-09 &    6   & $0\farcs6$ &  1.611  & 1.592  & 1.631 \\
2004-04-29 &   38   & $1\farcs2$ &  1.390  & 1.320  & 1.479 \\
2004-04-30 &   35   & $0\farcs9$ &  1.318  & 1.278  & 1.370 \\ \hline
\end{tabular}
\end{table}

  Imaging in the $Ks$-band was obtained in service mode with the SOFI (Son 
of ISAAC) imager on the ESO 3.5 m NTT at La Silla, Chile, during 5 
non-consecutive nights in April 2004.  We used the large field configuration,
providing an image scale of $0\farcs288$ pixel$^{-1}$ and a field size
of $5\arcmin\times5\arcmin$.  We made 265 exposures, each consisting 
of 10 co-adds for an integration time of 6$\times10$ = 60 s per exposure and
a total of 15900 s (4h25m).  A summary of the
SOFI observations is given in Table~\ref{tab:obslog}, which lists for each
night the number of exposures, the median seeing (measured on the
science images) and the airmass range.

  Our strategy for sky subtraction deserves a few explanatory notes: for 
uncrowded fields, sky subtraction in the near-IR is 
usually done by constructing a combined sky frame from several dithered 
exposures, using some rejection algorithm (e.g.\ median combination) to 
eliminate stars and other compact sources.  If the target fills a 
significant fraction of the field, dithering becomes impractical and 
separate sky frames must be obtained by nodding the 
telescope to an empty sky region. In our case, the GCs are unresolved at 
ground-based resolution and the field is sufficiently uncrowded that a
dithering approach would work well apart from the presence of NGC~4365. 
Although NGC~4365 does fill a significant 
fraction of the field of view, the profile is smooth and
over most of the field the surface brightness of the sky dominates completely
over the contribution from the galaxy. Thus, we opted \emph{not} to
obtain separate sky exposures and instead maximize the time spent on-target,
dithering the observations by applying a random offset within a 
$20\arcsec\times20\arcsec$ box before each new exposure. The penalty paid
for this approach is that we cannot perform proper sky subtraction near the 
very centre of the field where the contribution from NGC~4365 becomes
significant, but we considered this less of a problem since that region was
already covered by the $K$-band imaging of P02.

The standard SOFI calibration plan provides two standard stars per night,
but to ensure the best possible photometric calibration we requested that 4 
standard stars be observed on one photometric night together with some of
the science exposures.  These four stars (\object{S860-D}, \object{S791-C}, 
\object{S273-E} and 
\object{S870-T} in the list of Persson et al.\ \cite{per98}) were observed 
on 2004-04-30.

\subsection{SOFI $K$-band imaging: initial reductions and calibration}
\label{sec:sofical}

\begin{figure}
\includegraphics[width=85mm]{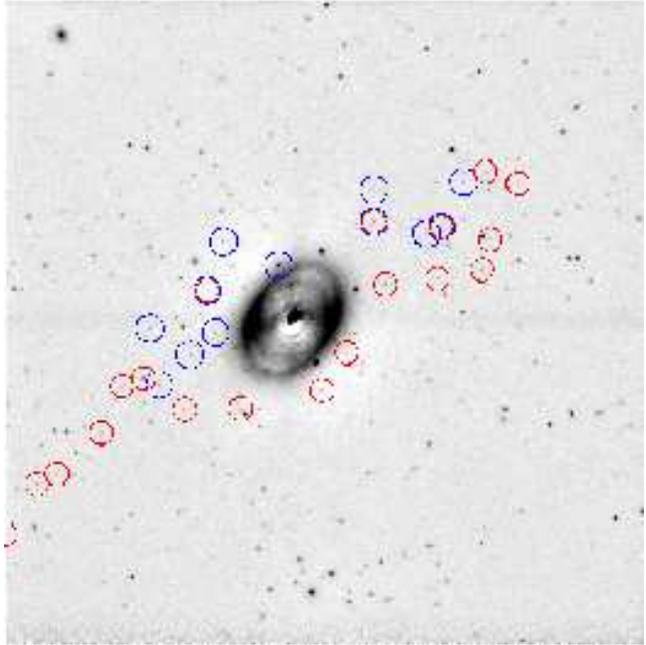}
\caption{SOFI K-band image of NGC~4365. Objects with spectroscopic data
  from Brodie et al.\ (\cite{bro05}) and Larsen et al.\ (\cite{lar03}) are 
  marked (red and blue, respectively, in the on-line edition). Sky background 
  and an 
  elliptical model of the galaxy (extending out to a semi-major axis of 
  $30\arcsec$) have been subtracted. The structure within the central 
  $\sim30\arcsec$ radius is entirely an artifact of poor background 
  subtraction; the galaxy light profile is very smooth all the way to the 
  centre (see text for details).}
\label{fig:mark_k}
\end{figure}

  First, the shifts between exposures were determined by measuring the
centroids of 3 bright, isolated stars in each image using the IMEXAMINE
task in IRAF\footnote{IRAF is distributed
by the National Optical Astronomical Observatories, which are operated by
the Association of Universities for Research in Astronomy, Inc.~under
contract with the National Science Foundation}.  We 
then constructed a model of the galaxy by flat-fielding the images with 
dome-flats, shifting and co-adding all the individual images, and fitting
an elliptical model of the galaxy light with the ELLIPSE and BMODEL tasks
in the STSDAS package in IRAF. Even after flat-fielding with dome flats,
non-uniformity of the background dominated over the galaxy light profile
at radii greater than $\sim30\arcsec$ from the centre, so we did
not attempt to model the galaxy light at larger radii. The elliptical
model was then shifted and subtracted from each of the raw frames.

  Next, each exposure (now with the central cusp of galaxy light removed, to
the extent possible) was reduced following the standard procedure: 
A ``running mean'' of the 4 exposures obtained closest in time (2 before 
and 2 after) was constructed 
for each frame by scaling each exposure to the same median value, rejecting 
any outliers (e.g.\ stars), and then forming the mean. For each exposure the 
corresponding running mean was then subtracted, and finally a flat-field 
correction was performed (using a dome-flat).

  The 265 exposures were then shifted and average combined.
Fig.~\ref{fig:mark_k} shows the combined image. The residuals of the galaxy
core are still clearly visible within the central $\sim30$ arcsec
radius, but over most of the field there are no remaining traces of the 
galaxy. The circles mark objects with spectroscopy from Paper I and Paper II.

  A photometric calibration was established using the four standard stars
observed on 2004-04-30. We used the PHOT task in the DAOPHOT package
(Stetson \cite{stet87})
in IRAF to perform aperture photometry within a radius of 8 pixels,
measuring the sky background in an annulus with inner and outer radii
of 20 and 30 pixels.  We assumed a transformation of the following form 
\begin{equation}
  Ks \, = \, ks \, + \, z_k \, - p_k x_k
  \label{eq:kcali}
\end{equation}
where $Ks$ and $ks$ are the standard and instrumental magnitudes and $p_k$ 
and $x_k$ are the extinction coefficient and airmass.  For the default PHOT 
zero-point of zmag=25 and assuming $p_k = 0.05$ mag (unit airmass)$^{-1}$,
we found $z_k=-2.667$ with an r.m.s. scatter of only 0.016 mag. We did not
attempt to perform an illumination correction, but each standard star
was observed at 5 positions across the detector and repeating the
zero-point determinations for each position separately we found all 
measurements to agree within 0.03 mag.  As a check we also derived $z_k$ 
for the other nights and found the zero-points for all nights to agree 
within 0.03 mag.  We thus consider the $K$-band photometric zero-points to 
be accurate within $\sim0.05$ mag across the SOFI field.

Photometry on the co-added science frame was done with a smaller ($r=4$)
aperture (see below). The zero-point offset between the $r=4$
photometry of the combined frame and the $r=8$ pixels standard calibration
of 2004-04-30 was determined in two steps: First an aperture correction
from $r=8$ to $r=4$ pixels was derived for the combined frame using a
standard curve-of-growth analysis as implemented in the MKAPFILE task in
DAOPHOT. Applying MKAPFILE to 13 bright, isolated objects distributed
across the SOFI field we found this correction to be 
$\Delta_{8\rightarrow4} = -0.255\pm0.019$ mag. 
As a check, we also
calculated the mean difference between photometry in $r=4$ and $r=8$
pixel apertures directly and got $\Delta_{8\rightarrow4} = -0.254\pm0.022$ 
mag.  
Next, the zero-point difference
between the combined frame and the night of 2004-04-30 was estimated by
combining all exposures from 2004-04-30 and comparing photometry in an
$r=8$ aperture with the combined frame. This difference was found to be
$\Delta_{2004-04-30\rightarrow{\rm all}} = -0.048\pm0.022$ mag. This
is somewhat larger than the scatter in the zero-points determined from the
standard star observations, but may be attributed to the fact that the
science data were obtained over a range of airmasses and transparency
variations may have occurred (we only required photometric conditions
for the night when standard stars were observed).  We thus
added an offset of $-0.303\pm0.03$ mag to the zero-point $z_k$ in 
Eq.~(\ref{eq:kcali}). 

\begin{figure}
\includegraphics[width=85mm]{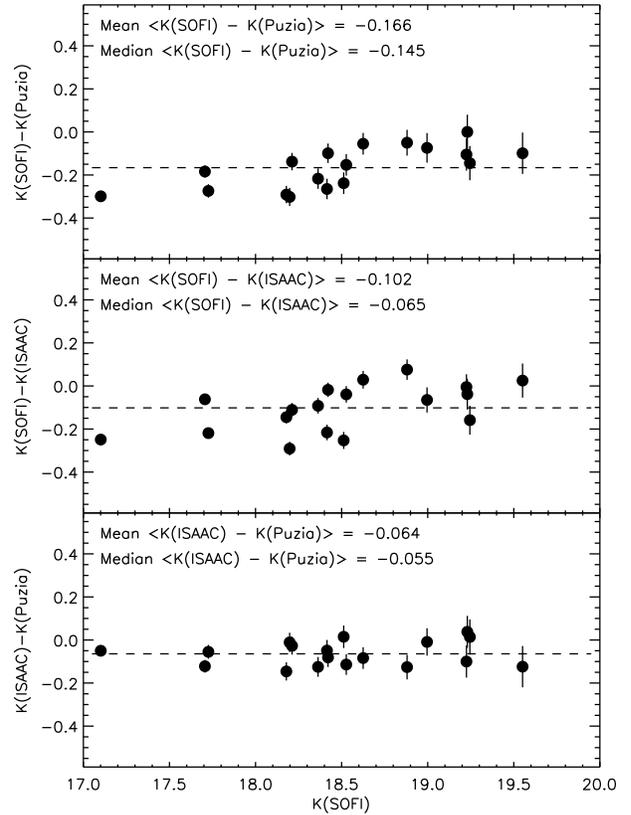}
\caption{Comparison of VLT/ISAAC and NTT/SOFI K magnitudes for isolated,
  bright objects. ISAAC magnitudes refer to measurements by us
  on a subset of the images used by P02, while ``Puzia'' refers
  to the P02 measurements.
}
\label{fig:kcmp}
\end{figure}

In the top panel of Fig.~\ref{fig:kcmp} we compare our $K$-band magnitudes 
for a set of isolated, bright objects in the NGC~4365 field with the ISAAC 
data of P02.  
The SOFI magnitudes are on average $-0.166\pm0.023$ mag brighter than those
of P02, with a hint of a systematic trend with magnitude in the sense 
that the absolute difference becomes smaller with fainter magnitudes.
As an independent check, we also downloaded a subset of the 
ISAAC data from the ESO-VLT archive and re-reduced them ourselves. The only
purpose of this exercise was to compare the
photometric calibrations for bright objects in the ISAAC and SOFI frames, so
we used a large aperture radius or $r=25$ pixels ($3\farcs7$) for both the 
GC and standard star observations and thereby eliminated the step of 
determining aperture corrections for the ISAAC data. The middle panel in 
Fig.~\ref{fig:kcmp} shows the comparison between our ISAAC reduction and 
the SOFI data. 
While the trend with magnitude persists, the mean offset with respect
to the SOFI data is reduced to $-0.102\pm0.026$ mag in our re-reduction of 
the ISAAC data. 
Finally, the bottom panel shows the comparison between
the P02 data and our ISAAC reduction. In this case there is no systematic
trend, and the mean offset ($-0.06$ mag) between the two reductions is 
consistent with the differences in the mean offsets with respect to 
the SOFI data.

 The comparison between the various $K$-band measurements illustrates
the difficulty of accurately calibrating near-IR data due to issues
such as uncertain photometric zero-points, aperture corrections, and
the bright sky background.
Even for the brightest GCs in the field, the count rate from the
object is generally less than 1\% of the sky background. Variations
in the background itself of a similar magnitude occur over timescales of a few
minutes, so accurate background subtraction is a critical issue which
might be compromised e.g. by non-linearities in the detector.
We have argued that our photometric zero-points are likely accurate 
to better than 0.03 mag, with an additional 0.03 mag uncertainty on the 
aperture corrections.  These could be underestimates, however.  
P02 argue that their photometric calibration 
is accurate to $\la0.03$ mag, but given the $\sim0.06$ mag offset in the 
bottom panel of Fig.~\ref{fig:kcmp} it seems likely that this might also be 
somewhat underestimated.  We believe that these differences simply reflect 
the inherent difficulty of accurately calibrating near-infrared photometry, 
due in large part to the less well-behaved infrared detectors 
(compared to CCDs).  Like our SOFI data, the ISAAC data were also obtained 
over several nights, and it is not \emph{a priori} clear to us that the
calibration of one dataset should be considered more reliable than the other.
Rather, we infer from this comparison that the true calibration uncertainty
on either dataset is probably on the order of 0.1 mag.  

  It is hard to tell whether the trend with magnitude seen in the two top 
panels of Fig.~\ref{fig:kcmp} is caused by the ISAAC or the SOFI data. The
overall sense of the offset between the two datasets makes it most likely
that the SOFI $K$ magnitudes are slightly too bright, which would tend to
make the \vk\ colours systematically too red. The implications are discussed 
in more detail below (\S\ref{sec:thecase}).

\subsection{FORS1 $V$ and $I$-band imaging}
\label{sec:fors1}

\begin{table}
\caption{\label{tab:forslog}Log of FORS1 observations}
\begin{tabular}{lcccc} \hline
          & Date       & T$_{\rm exp}$  &  Seeing      & Airmass \\ \hline
 $V$-band & 2003-01-01 & $3\times140$ s &  $0\farcs80$ & 1.50 \\
 $I$-band & 2002-12-30 & $3\times190$ s &  $0\farcs55$ & 1.35 \\ \hline
\end{tabular}
\end{table}

  The SOFI data were combined with $V$ and $I$ imaging from the
FORS1 instrument on the ESO VLT (Table~\ref{tab:forslog}). These data
were originally obtained as pre-imaging for a (never completed) spectroscopic 
study 
(70.B-0289, P.I.\ Kuntschner) and are not very deep (3$\times$140 s 
and 3$\times190$ s in $V$ and $I$) but are well matched to our $K$ data.
The FORS1 image scale was $0\farcs2$ pixel$^{-1}$ and the field of view
$6\farcm8\times6\farcm8$, completely including the SOFI field.

  Initial processing of the CCD images (bias subtraction, flat-fielding)
was done using standard IRAF tools and the flat-fields (sky flats) and 
bias exposures provided by the ESO archive.
For the photometric calibration we used observations of two standard
fields (PG0231 and Rubin 152 from Landolt \cite{lan92}), observed on
2002-12-29, 2003-01-01 and 2003-01-06. As the standard star exposures were
strongly defocused (in order to avoid saturation), we used a large
($r=20$ pixels) aperture for the standard star photometry.
The transformations from FORS1
instrumental to standard Johnson-Cousins $V$ and $I$ magnitudes were 
assumed to be of the form (using the same notation as in Eq.~\ref{eq:kcali}):
\begin{equation}
  V \, = \, v_e + z_v + c_v (v_e-i_e) 
\end{equation}
and
\begin{equation}
  I \, = \, i_e + z_i + c_i (v_e-i_e) 
\end{equation}
where $v_e = v - k_v x_v$ and $i_e = i - k_i x_i$.
We found zero-points of $z_v=2.510\pm0.004$ mag and $z_i=1.535\pm0.008$
mag and colour terms $c_v=0.053\pm0.008$ and $c_i=-0.027\pm0.016$, 
assuming $k_v=0.10$ and $k_i=0.04$ and the default PHOT zero-point
of zmag=25. We found it necessary to eliminate the bluest star in
the PG0231 field ($\vi =-0.534$), or otherwise a simple linear colour
term did not give a good fit. 
The remaining stars had colours
between $\vi=-0.145$ and 1.951, adequately bracketing the typical
colours of GCs.  
Aperture corrections
from the $r=20$ to $r=4$ pixels apertures used for photometry on the
science frames were again determined with the
MKAPFILE task, and were found to be 
$\Delta_{\rm20\rightarrow4}(V) = -0.336\pm0.015$ mag and
$\Delta_{\rm20\rightarrow4}(I) = -0.216\pm0.008$ mag.  Our formal errors
on the photometric calibration are thus smaller than 0.02 mag in both $V$ 
and $I$.  Note the smaller aperture correction in $I$, due to the 
better seeing. 

\subsection{$VIK$ Photometry of the NGC~4365 GC system}

\begin{figure}
\includegraphics[width=85mm]{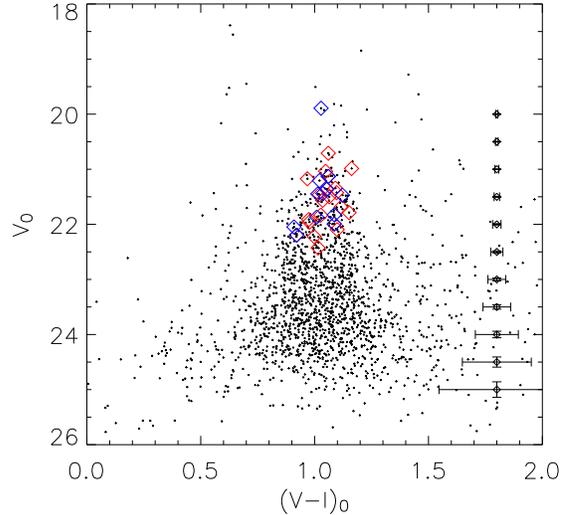}
\caption{FORS1 \vi, V colour-magnitude diagram for objects in NGC~4365.
 Objects with spectroscopy from Larsen et al.\ (\cite{lar03}) and
 Brodie et al.\ (\cite{bro05}) are shown with diamond symbols
 (blue and red in the on-line edition).
}
\label{fig:cmd_vi}
\end{figure}

\begin{figure}
\includegraphics[width=85mm]{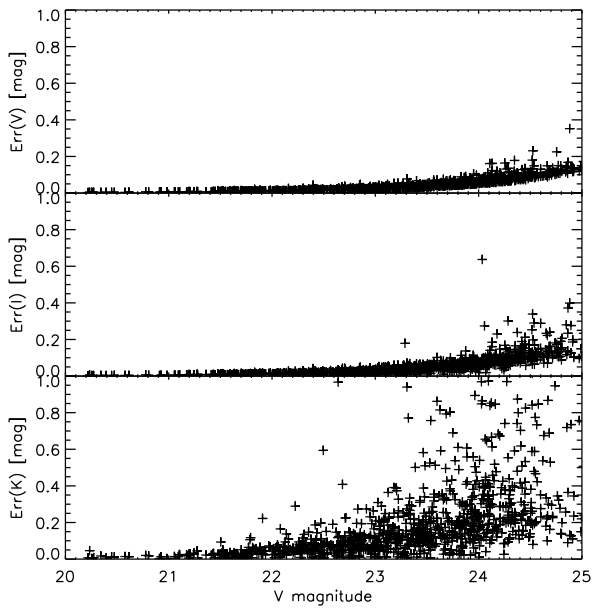}
\caption{Photometric errors in the $V$, $I$ and $K$ bands as a function
 of $V$ magnitude.
}
\label{fig:perr}
\end{figure}

\begin{figure}
\includegraphics[width=85mm]{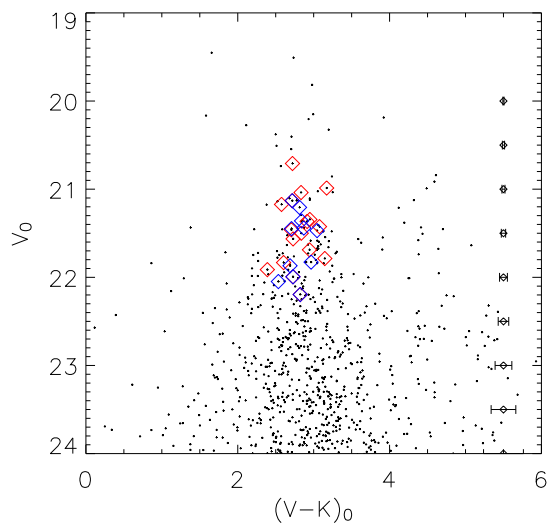}
\caption{SOFI/FORS1 \vk, V colour-magnitude diagram for objects in NGC~4365.
 Objects with spectroscopy from Larsen et al.\ (\cite{lar03}) and
 Brodie et al.\ (\cite{bro05}) are shown with diamond
 symbols (blue and red in the on-line edition).
}
\label{fig:cmd_vk}
\end{figure}

  The FORS1 $V$ and $I$ images were registered to a common reference
system and the galaxy light profile was modelled and subtracted using the 
ELLIPSE and BMODEL tasks in STSDAS. Objects were then
detected in the $V$ band image using the DAOFIND task in DAOPHOT, adopting 
a 4$\sigma$ threshold above the background noise estimated in the 
\emph{outer} parts of the image. Closer to the centre where the
background noise is higher, this led to a large number of false detections 
and the initial object list was cleaned by only including objects 
which represented a 5$\sigma$ or better detection above the \emph{local} 
background noise, measured in a small annulus around each object.

  Photometry was then obtained for each source using the PHOT task
with an aperture radius of 4 pixels and calibrated as described above.
Fig.~\ref{fig:cmd_vi} shows the resulting $V$ vs.\ \vi\ colour-magnitude 
diagram, corrected for Galactic foreground extinction (Table~\ref{tab:gdat}). 
The GC population is
the dominant feature in the plot, with $(\vi)_0$ colours between
$\approx0.8$ and $\approx1.2$ and extends over the entire magnitude range 
between
$20 \la V \la 24$. In L2001 we found the turn-over of the GC luminosity
function to be at $V=24.22^{+0.13}_{-0.14}$ 
so here we are just barely reaching down to the
turn-over. The GCs with spectroscopy from Papers I and II are marked
with diamond symbols and sample the GC colour range quite well, though 
perhaps with some bias towards intermediate and red colours (where most 
candidate intermediate-age GCs would be expected).

\begin{table*}
\caption{\label{tab:vik}
  FORS1 and SOFI photometry. Magnitudes listed here are not corrected for
  foreground extinction.  The FWHM is measured by ISHAPE on
  the FORS1 $V$-band image and is the intrinsic FWHM of 
  the object in pixel units, corrected for the FORS1 PSF. The
  full version of this table is available online or upon request
  from the authors.}
\begin{tabular}{lccccccccccc} \hline
 ID & \multicolumn{2}{c}{X,Y (FORS1)} & RA (2000.0) & Decl (2000.0) & $V$ & $\sigma V$ & $I$ & $\sigma I$ & $Ks$ & $\sigma Ks$ & FWHM \\ \hline
131 & 1482.6 & 244.3 & 12:24:21.76 & 7:16:26.6 & 21.754 & 0.009 & 19.819 & 0.005 & 17.791 & 0.013 & 0.33 \\
132 & 1767.5 & 243.8 & 12:24:17.93 & 7:16:26.3 & 24.277 & 0.067 & 23.295 & 0.087 & $\ldots$ & $\ldots$ & 1.58 \\
133 & 1386.6 & 246.3 & 12:24:23.05 & 7:16:27.1 & 24.388 & 0.075 & 23.251 & 0.082 & 20.353 & 0.157 & 0.19 \\
\hline
\end{tabular}
\end{table*}

  A coordinate transformation between the FORS1 and SOFI data was
established by identifying 21 objects common to both datasets
and measuring their $(x,y)$ coordinates with the IMEXAMINE task in IRAF. The 
transformation was then computed using the GEOMAP task in the IRAF IMMATCH 
package. The r.m.s. residuals around the transformation were about 0.2 
pixels in both $x$ and $y$. The FORS1 object
list was transformed to the SOFI frames and $Ks$ photometry was obtained
for each object using an aperture radius of 4 pixels. The central
$\sim30\arcsec$ radius region of the SOFI image, where the background
subtraction is poor, was masked out.  The instrumental
magnitudes were calibrated to standard $Ks$ magnitudes as described 
in \S\ref{sec:sofical}. Table~\ref{tab:vik} lists a few rows from
the combined photometry data file. The full version of the table is
available electronically. 
The photometry in the table is not corrected
for foreground reddening but for our subsequent analysis we use
the Schlegel et al.\ (\cite{sch98}) value of $A_B=0.091$ mag and
the reddening law in Cardelli et al.\ (\cite{car89}), i.e.\
$A_B = 1.337 \, A_V$, $A_I = 0.563 \, A_V$ and $A_K = 0.114 \, A_V$.
Note that Table 3 in Cardelli et al.\ gives the $A_I/A_V$ ratio for
the Johnson $I$ filter centered at 900 nm, whereas our photometry is
calibrated to the Kron-Cousins $I$ band which is centered at
about 825 nm (Landolt \cite{lan83}). We have adopted the corresponding
value for the $A_I/A_V$ ratio, using the general expressions for
$A(\lambda)/A_V$ given in Cardelli et al.

  Because the objects were detected in the FORS1 images the detection
completeness is not formally dependent on the $K$-band data. However, at 
a given magnitude the photometric errors are always dominated by the 
$K$-band data (Fig.~\ref{fig:perr}) and rise rapidly below $V=23$ 
(corresponding to $K\approx20-20.5$ for typical GC colours).

  The $V$ vs.\ \vk\ colour-magnitude diagram is shown in 
Fig.~\ref{fig:cmd_vk}. As in Fig.~\ref{fig:cmd_vi}, objects with
spectroscopy are indicated with diamond symbols. A few of these fall
within the central masked-out region or outside of the SOFI field and
hence are not plotted. A total of 25 objects have both $VIK$ photometry
and spectroscopy (11 from Paper I and 17 from Paper II, with three
objects in common between the two papers) and again the \vk\ colour range 
is well sampled by the combined spectroscopic samples.

\subsection{HST WFPC2 imaging}
\label{sec:wfpc2cal}

\begin{figure}
\includegraphics[width=85mm]{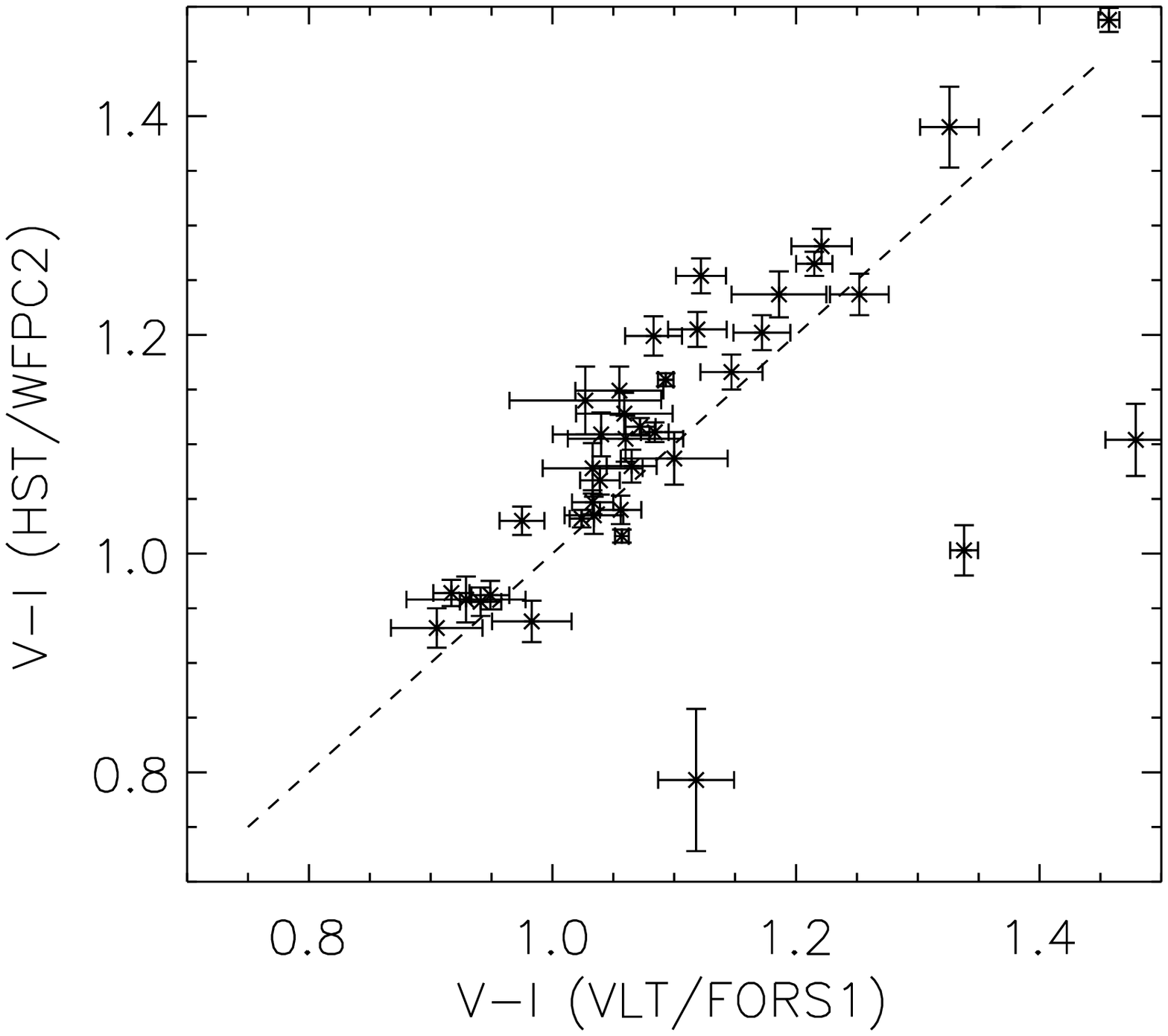}
\caption{Comparison of VLT/FORS1 and HST/WFPC2 \vi\ colours for objects
  with $20<V<23$. The mean and median difference are
  $\langle (\vi)_{\rm FORS1} - (\vi)_{\rm WFPC2}\rangle = -0.017\pm0.024$ mag 
  and
  $\langle (\vi)_{\rm FORS1} - (\vi)_{\rm WFPC2}\rangle_{\rm Med} = -0.030$ mag.
}
\label{fig:vivi}
\end{figure}

\begin{figure}
\includegraphics[width=85mm]{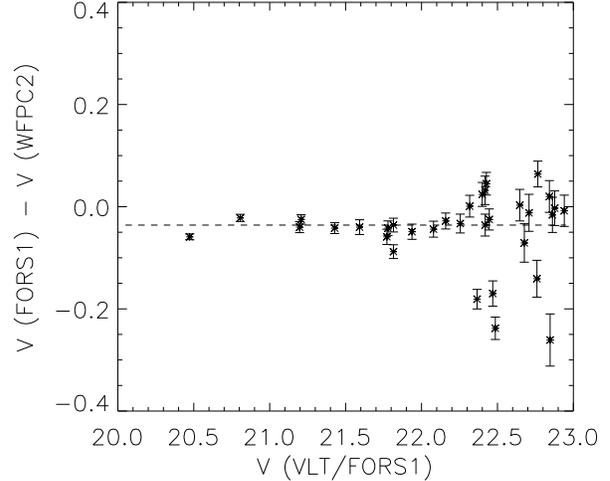}
\caption{Comparison of VLT/FORS1 and HST/WFPC2 V magnitudes for objects
  with $20<V<23$. The mean and median difference are
  $\langle V_{\rm FORS1} - V_{\rm WFPC2}\rangle = -0.045\pm0.013$ mag.
  $\langle V_{\rm FORS1} - V_{\rm WFPC2}\rangle_{\rm Med} = -0.036$ mag.
}
\label{fig:vv}
\end{figure}

  In addition to the SOFI and FORS1 data, we use data taken with the
Wide Field Camera 2 (WFPC2) and Advanced Camera for Surveys (ACS) on board
the Hubble Space Telescope (HST). The homogeneity and stability of HST 
data allow us to carry out a robust comparison of GCs in NGC~4365 with
three other large Virgo ellipticals with rich GC systems: NGC~4406 (M86), 
NGC~4486 (M87) and NGC~4649 (M60).  

  The WFPC2 data are the same data that were used in L2001 and we refer to 
that paper for details concerning the reductions. Briefly, the data sets 
consist of deep (1 orbit or $\sim$2400 s) CR-SPLIT exposures in each of 
the F555W ($V$) and F814W ($I$) bands.  For NGC~4365 we used the two pointings 
from programmes 5920 and 6554 (P.I.\ Brodie).  A very large number of pointings 
are available for NGC~4486 but we restricted our analysis to the two 
closest to the centre (from programmes 5477 and 6844, P.I.\ Macchetto) 
labelled NGC~4486 and NGC~4486-O1 in L2001. For each of NGC~4406 and 
NGC~4649 only one pointing was available. 

  Most of the exposures are accurately aligned at the sub-pixel 
level but, as noted in L2001
shifts are present between the exposures
in the central NGC~4365 pointing. A re-analysis of this data showed
that the resampling of the data prior to the co-addition demanded 
more relaxed cosmic-ray rejection parameters than those used in 
L2001 in order to avoid artifacts near the centre of bright sources. 

  We have also revised the aperture corrections for the WFPC2 data by
directly measuring the difference between photometry in an $r=5$ pixels
($0\farcs5$) aperture and our $r=2$ pixels apertures used for \vi\ colours.
We find some differences from one dataset to another, ranging between
$\Delta_{\rm5\rightarrow2}(\vi) = 0.02$ mag and
$\Delta_{\rm5\rightarrow2}(\vi) = 0.07$ mag. In L2001 we used a constant
aperture correction for \vi\ of 0.026 mag.  For the $r=3$ apertures used
for $V$ magnitudes we find
$\Delta_{\rm5\rightarrow3}(V) = -0.07$ mag to
$\Delta_{\rm5\rightarrow3}(V) = -0.11$ mag. The Holtzman et al.\ (\cite{hol95})
photometric calibration of WFPC2 includes an implicit $-0.1$ mag 
correction from $r=0\farcs5$ to infinity, although this may be an 
underestimate for extended sources such as GCs.  In L2001 we added a
$-0.07$ mag correction to account for the fact that some fraction of the 
light from a typical GC will extend beyond the $0\farcs5$ aperture, but this 
correction was actually derived for a smaller distance than that of the 
four galaxies studied here and is thus likely to be an overestimate (as 
noted in L2001). For our present purpose a small uncertainty on the WFPC2 
$V$ magnitudes is unimportant, as we are mainly interested in the \vi\ 
colour distributions.  Thus, we do not include this additional zero-point 
correction here. 

The different aperture corrections lead to shifts of a few times 0.01 mag
in the \vi\ colour distributions with respect to L2001, while the improved
CR-rejection parameters for NGC~4365 cause a systematic shift of almost 
0.05 mag in \vi\ as well as a decreased scatter. This shift 
is clearly seen when comparing the new \vi\ colour distribution
(Fig.~\ref{fig:vihist}) with Figure 4
in L2001.  We have verified that no residuals are
present near the centres of bright objects in our new reduction of the
NGC~4365 WFPC2 data and our WFPC2 \vi\ colours for NGC~4365 agree with
those of P02 (derived from the same data) within a mean difference 
of $\Delta (\vi)_{\rm 5920-P02} = 0.019\pm0.010$ mag.  Since we used similar 
apertures to those of P02 (r=2 pixels) for the \vi\ colours, the source
of the remaining difference is unclear, but may be related to the fact that 
we used DAOPHOT for the photometry whereas P02 used SExtractor. Somewhat
more worrisome is a systematic difference of 
$\Delta V_{\rm 5920-P02} =  -0.12\pm0.01$ mag between our WFPC2 $V$ 
magnitudes and those of P02. In order to be less sensitive to uncertainties
in the aperture corrections, we used an $r=3$ pixels aperture for our
$V$ magnitudes, while P02 used the same $r=2$ aperture for the $V$
magnitudes as for \vi\ colours. As a check, we derived aperture corrections
between $r=2$ and $r=5$ pixels ($0\farcs5$) for our photometry and
compared with the aperture corrections given by P02. Our aperture corrections 
($-0.20$ mag in F555W) are about $-0.06$ mag larger than those in P02 
($-0.14$ mag), leaving an additional $0.06\pm0.01$ mag unaccounted for. 

  In Figs.~\ref{fig:vivi} and \ref{fig:vv} we compare our WFPC2 $V$ and $I$
photometry
with the FORS1 data. There is excellent agreement between the FORS1 and 
WFPC2 \vi\ colours, with
a mean and median systematic difference of $-0.017\pm0.024$ mag and
$-0.03$ mag, consistent with our estimated calibration uncertainties. The 
FORS1 and WFPC2 $V$ magnitudes also agree
very well, with a mean and median difference of $-0.045\pm0.015$ mag
and $-0.036$ mag, the FORS1 magnitudes being slightly brighter. This difference
could be partly due to the extended nature of the GCs. Note, however,
that the offset of our FORS1 photometry with respect to the P02 WFPC2 data
is larger, about $-0.16$ mag. Incidentally, this is similar to the 
difference between the two sets of $K$-band measurements, making
our \vk\ colours similar to those of P02.

An additional WFPC2 F555W and F814W
dataset is available for NGC~4365 (programme 5454). We have also downloaded
and reduced this dataset, and find excellent agreement (within
the errors) with the deeper data used here. Specifically, the mean colour 
and magnitude differences are $\Delta (\vi)_{5920-5454} = -0.024\pm0.010$ mag 
and $\Delta V_{5920-5454} = -0.017\pm0.05$ mag. 

\subsection{HST ACS imaging}

  The ACS data were all obtained as part of the ACS Virgo Cluster Survey
(C{\^o}t{\'e} et al.\ \cite{cot04}) and use the F850LP ($\sim$Sloan $z$)
and F475W ($\sim$Sloan $g$) filters. The wider separation in central 
wavelength of these filters compared to \vi\ allows better resolution of 
features in the GC colour distributions. The exposure times in F850LP and
F475W were 1210 and 750 seconds, split into 3 and 2 exposures for
cosmic-ray rejection. The data were obtained with the wide field channel
on ACS, which has a pixel scale of $0\farcs050$ pixel$^{-1}$ and a 
field of view of $203\arcsec\times203\arcsec$.

  ACS images of NGC~4365, NGC~4406, NGC~4486 and NGC~4649 were downloaded
from the archive at the \emph{Space Telescope European Coordinating
Facility} (ST-ECF). The pipeline processed images were combined with
the MULTIDRIZZLE task (Koekemoer et al.\ \cite{koek02}) in the 
STSDAS.DITHER package which removes the significant geometric distortion 
in ACS images and normalises the images to count rate.

  Objects were detected with SExtractor (Bertin \& Arnouts \cite{ba96}) 
and photometry was obtained with PHOT, using an aperture radius of 
5 pixels. Aperture corrections were derived by computing the mean
difference between the $r=5$ pixels aperture and a reference $r=20$
aperture for objects with $20 < m_{\rm F850LP} < 22$. We found mean
aperture corrections of 
$\Delta m_{\rm 20\rightarrow5}({\rm F475W})=-0.13$ mag and
$\Delta m_{\rm 20\rightarrow5}({\rm F850LP})=-0.23$ mag to be
adequate for all galaxies (within 0.02 mag).
The photometry was calibrated to AB magnitudes using the zero-points on 
the December 16, 2004 version of the ACS Web 
Page\footnote{http://www.stsci.edu/hst/acs/analysis/zeropoints}
(Z$_{AB}$(F475W)=26.068 mag, Z$_{AB}$(F850LP)=24.862 mag).
The ACS F850LP and F475W filters do not exactly match the Sloan $g$ and
$z$ filters but for this paper we only use the ACS data to compare the
colour distributions of GCs in the four galaxies in a relative sense.
In order to avoid the regions near the centre of the galaxies, objects 
with a background level higher than 0.75 counts s$^{-1}$ pixel$^{-1}$
in F850LP were excluded from the analysis.

\section{The GC system of NGC~4365 - peculiar colour distribution?}
\label{sec:peccol}

\begin{table*}
\caption{\label{tab:gdat}
 Data for the galaxies. Classifications are from the
 NASA/IPAC Extragalactic Database (NED). The \vi\ colours and $m-M$ are 
 from Tonry et al.\ (\cite{ton01}), $m_B$ and $B-V$ from RC3.
 $N_{\rm GC}$ (ACS) is the number of GC candidates in the ACS frames
 with m$_{\rm F850LP}<23.5$ and 0.7$<$F475W-F850LP$<$1.7}
\begin{tabular}{lcccccccc} \\ \hline
         & Type & $(\bv)_0$  &  $(\vi)_0$  & $m-M$ & $A_B$ & $m_B$ &  $M_B$  & $N_{\rm GC}$ (ACS) \\ \hline
NGC~4365 & E3    & 0.95 & 1.222 & 31.55 & 0.091 & 10.50 & $-21.1$ &  476 \\
NGC~4406 & S0/E3 & 0.90 & 1.167 & 31.17 & 0.128 &  9.80 & $-21.5$ &  267 \\
NGC~4486 & E0-1  & 0.93 & 1.244 & 31.03 & 0.096 &  9.50 & $-21.6$ & 1124 \\
NGC~4649 & E2    & 0.95 & 1.232 & 31.13 & 0.114 &  9.80 & $-21.4$ &  533 \\ \hline
\end{tabular}
\end{table*}

\begin{figure}
\includegraphics[width=85mm]{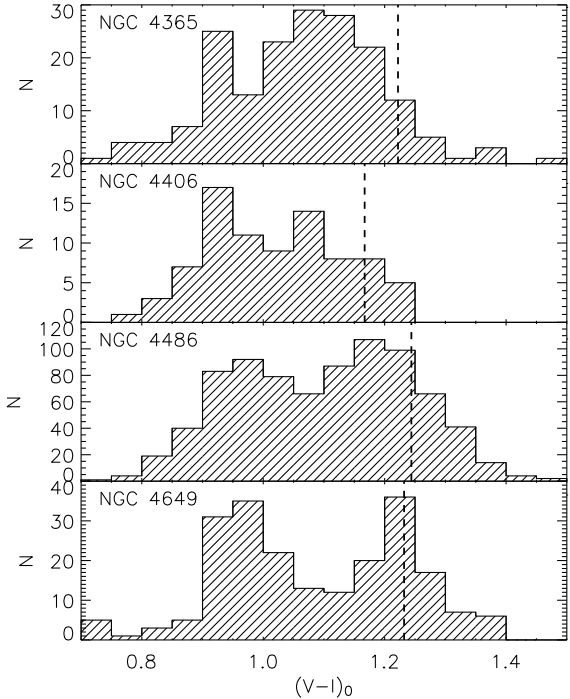}
\caption{Histograms of the \vi\ colour distributions for GC candidates in
  NGC~4365, NGC~4406,
  NGC~4486 and NGC~4649, based on HST/WFPC2 imaging from 
  Larsen et al.\ (\cite{lar01}). Corrections for foreground reddening
  from Schlegel et al.\ (\cite{sch98}) have been applied. The vertical
  dashed lines indicate the integrated galaxy colours from 
  Tonry et al.\ (\cite{ton01}). 
}
\label{fig:vihist}
\end{figure}

\begin{figure}
\includegraphics[width=85mm]{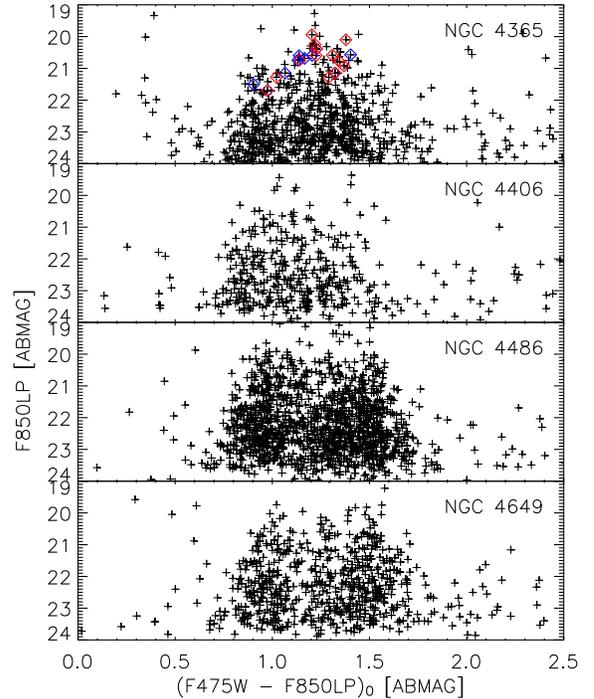}
\caption{F850LP vs. F475W--F850LP colour-magnitude diagrams for GCs in
  NGC~4365, NGC~4406, NGC~4486 and NGC~4649, based on HST/ACS imaging from 
  the Virgo Cluster Survey (C{\^o}t{\'e} et al.\ \cite{cot04}). Corrections 
  for foreground reddening from Schlegel et al.\ (\cite{sch98}) have 
  been applied. As in Fig.~\ref{fig:cmd_vi} and \ref{fig:cmd_vk}, objects
  with spectroscopy from Paper I and II are indicated.
}
\label{fig:gzcmd}
\end{figure}

\begin{figure}
\includegraphics[width=85mm]{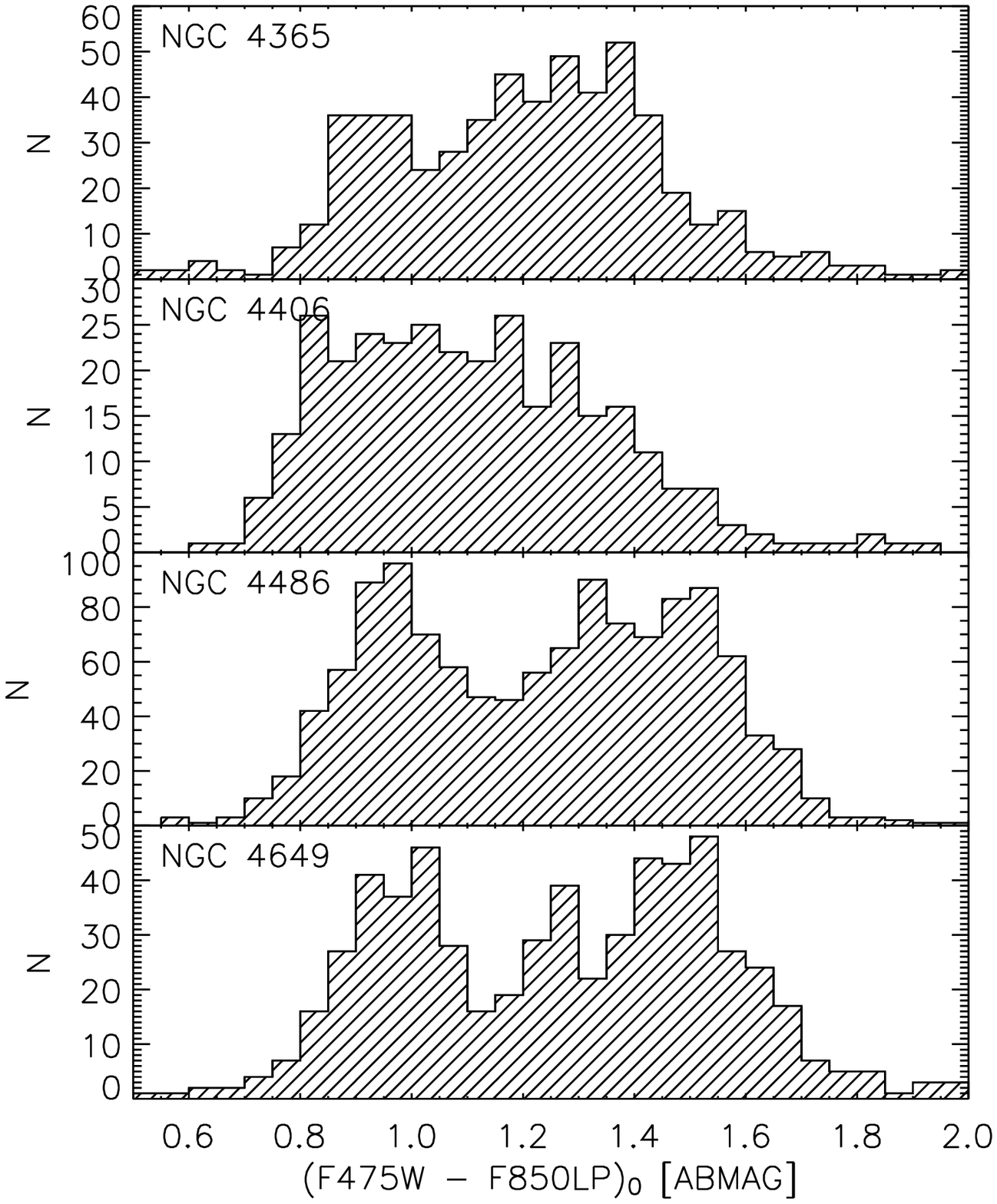}
\caption{Histograms of the F475W--F850LP colour distributions for 
  NGC~4365, NGC~4406, NGC~4486 and NGC~4649, based on HST/ACS imaging from the
  Virgo Cluster Survey (C{\^o}t{\'e} et al.\ \cite{cot04}). Corrections 
  for foreground reddening from Schlegel et al.\ (\cite{sch98}) have 
  been applied. Bi- or even tri-modality is evident in NGC~4365, NGC~4486
  and NGC~4649, but the red peak is suppressed and/or shifted towards 
  the blue in NGC~4365.
}
\label{fig:gzhist}
\end{figure}

We begin our analysis by comparing the NGC~4365 GC system with those of
NGC~4406, NGC~4486 and NGC~4649, chosen as representative comparison cases.
A complete discussion of the general properties of GC systems in different 
galaxies is not intended here and we refer to previous work by many 
authors (Gebhardt \& Kissler-Patig \cite{gk99}; Kundu \& Whitmore \cite{kw01}; 
L2001). We concentrate on aspects relevant to the comparison with
NGC~4365.

Some basic properties of the four
galaxies are listed in Table~\ref{tab:gdat}. They are all members of
the Virgo cluster and have comparable absolute luminosities although
NGC~4365 is the fainter of the four if we adopt the distance moduli
of Tonry et al.\ (\cite{ton01}). The last column, $N_{\rm GC}$, lists
the number of GC candidates brighter than $m_{\rm F850LP}=23.5$ and
with $0.7<$F475$-$F850LP$<1.7$ detected in the ACS images. This does
not include any completeness corrections and is by
no means intended to give a complete estimate of the number of GCs 
in each galaxy, but merely serves to illustrate relative differences in
the richness of the GC systems within the central few arcmin.  
Of particular interest here are the integrated colours of the four galaxies. 
The \vi\ colours are taken from Tonry et al.\ (\cite{ton01}) and are measured 
within an aperture radius of $\sim30\arcsec$ (except for NGC~4486 where 
a radius of $53\arcsec$ was used), while the $B-V$ colours are total
asymptotic colours from RC3 (de Vaucouleurs et al.\ \cite{rc3}). NGC~4365, 
NGC~4486 and NGC~4649 have similar colours, while NGC~4406 is bluer in
both \bv\ and \vi. According to the \emph{NASA/IPAC Extragalactic Database}
(NED) all galaxies except NGC~4406 are classified as ellipticals, while 
NGC~4406 is listed as S0/E3.

Fig.~\ref{fig:vihist} shows the \vi\ colour distributions of GC candidates
with $20<V<23.5$ in the four galaxies.  Contrary to our previous finding
(L2001), the \vi\ histogram for NGC~4365 does show a 
clearly bimodal structure in our re-reduced data with a narrow blue peak at 
$(\vi)_0\approx0.95$ and a broader peak at $(\vi)_0\approx1.1$ (read off by 
eye from the figure).  NGC~4486 and NGC~4649 also show bimodal colour 
distributions with blue peaks at about the same colour as in NGC~4365 or 
perhaps slightly redder. The red peaks in these two galaxies are significantly 
redder than in NGC~4365. In NGC~4406 the evidence for bimodality is weaker
although there is still a hint of a blue peak at about the same colour as
in the other galaxies. 

The vertical dashed lines in Fig.~\ref{fig:vihist} indicate the integrated
galaxy \vi\ colours from Table~\ref{tab:gdat}.  The mean colour of the red
GCs in NGC~4365 is shifted towards the blue with respect to the galaxy
light, although the tail of the colour distribution includes some
objects with colours as red as the galaxy. 
This is quite different from
the situation in NGC~4486 and NGC~4649 where the red GCs have mean
colours which are more similar to the galaxy light.
Though the overall GC colour distribution in 
NGC~4406 appears bluer than in NGC~4486 and NGC~4649, this also seems
consistent with the bluer colour of the galaxy itself.

In Figs.~\ref{fig:gzcmd} and \ref{fig:gzhist} we show the ACS
colour-magnitude diagrams and corresponding colour distributions
for objects with m(F850LP)$<$23.5. The colour
distributions mimick those seen in Fig.~\ref{fig:vihist} with clear
bimodality in NGC~4365, NGC~4486 and NGC~4649.
Again, the blue peaks are at about the same colour (perhaps
slightly bluer in NGC~4365), and the red GCs in NGC~4365 appear more
strongly weighted towards intermediate colours compared to NGC~4486 and 
NGC~4649. As in Fig.~\ref{fig:vihist}, distinct peaks are difficult to 
identify in NGC~4406, but there is a relative paucity of very red clusters 
compared to NGC~4486 and NGC~4649. 
Brodie et al.\ (\cite{bro05}) suggested that the \gz\ colour distribution 
in NGC~4365 may actually be \emph{tri}-modal; we cannot confirm that here 
but the red peak does appear quite broad and there are hints that the GCs 
with intermediate colours are concentrated closer towards the centre of
NGC~4365 (\S\ref{sec:radial}).  Based on the \gz\ histograms,
three peaks might also be present in NGC~4486 (as suggested
by Lee \& Geisler \cite{lg93}) and NGC~4649, and
trimodality has even been claimed for NGC~4406 (Rhode \& Zepf \cite{rz04}).

Another interesting feature to note from the colour-magnitude diagrams
is that the brightest intermediate-colour and red GCs in NGC~4365 are
$\sim1$ mag brighter than the brightest blue clusters. The brightest red GCs 
have $m_{\rm F850LP}\approx20$, while the brightest blue ones have
$m_{\rm F850LP}\approx21$.  Since the number of clusters per colour bin
is roughly similar for blue and intermediate colours, this cannot simply be 
a size-of-sample effect.  In the other galaxies, the brightest blue and 
red GCs are comparable in magnitude and few if any clusters are brighter
than $m_{\rm F850LP}\approx20$. Considering the somewhat greater distance
of NGC~4365, the brightest red GCs in this galaxy may be slightly
over-luminous compared to those in NGC~4486, NGC~4649 and NGC~4406.
This difference would be expected if a significant 
population of intermediate-age GCs were present in NGC~4365, due to the 
increase in mass-to-light ratio with age. Indeed, this effect has been used 
in several previous studies in attempts to identify candidate intermediate-age 
GC populations in suspected merger remnants 
(Whitmore et al.\ \cite{whit97}; Brown et al.\ \cite{bro00}).
Of course, an alternative possibility is that the \emph{mass} distribution of
the red GCs in NGC~4365 simply extends to higher masses.

At any rate, the nature of the brightest GCs in galaxies remains unclear. 
The brightest GC in the Milky Way, $\omega$ Cen, displays a range of 
peculiarities including a large flattening, a wide metallicity distribution 
and at least two distinct stellar populations (Piotto et al.\ \cite{pio04}).
A similar wide metallicity distribution has been noted for G1, one of 
the brightest GCs in M31 (Meylan et al.\ \cite{mey01}). In the giant
Virgo elliptical \object{NGC~4636}, Dirsch et al.\ (\cite{dir05}) found a 
larger proportion of intermediate-color GCs at the bright end of the luminosity
distribution than at fainter magnitudes and Mieske et al.\ (\cite{mie02})
suggested a smooth transition from the brightest GCs in NGC~1399 to 
Ultra Compact Objects.  In NGC~1399, Richtler et al.\ (\cite{tom05}) have
noted 8 bright globular cluster-like objects with peculiar morphological
properties.  Thus, the very brightest objects in NGC~4365 (and other galaxies) 
might not be directly related to the normal, fainter GCs.

\section{The case for an intermediate-age population?}
\label{sec:thecase}

\subsection{SOFI/FORS1 $VIK$ photometry}
\label{sec:sofi}

\begin{figure}
\includegraphics[width=85mm]{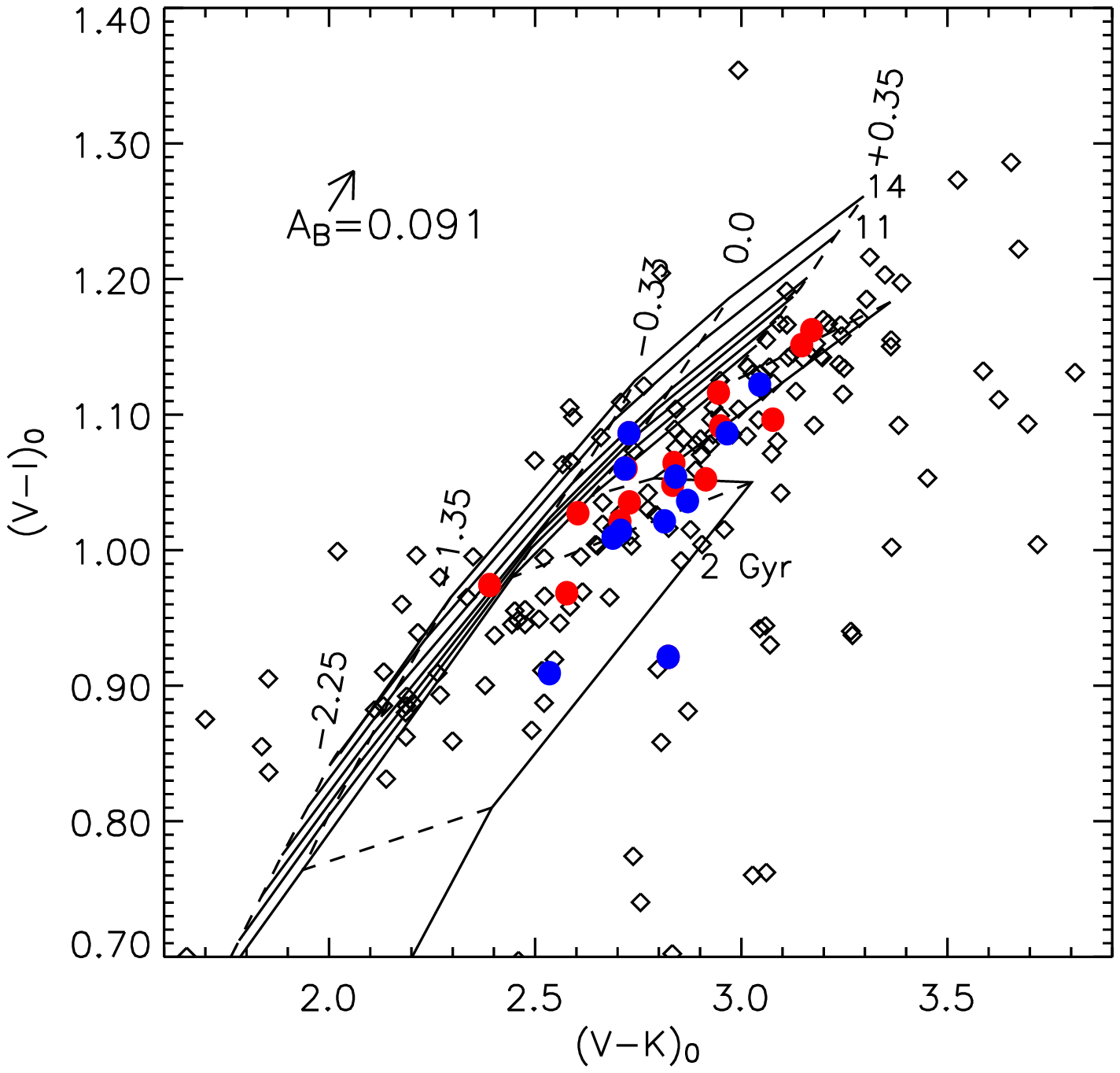}
\caption{(\vk, \vi) diagram for objects in NGC~4365 
  and SSP models from Maraston (2004). Model grids are shown for
  ages of 2, 3, 4, 5, 6, 8, 11 and 14 Gyrs (solid lines) and metallicities
  $[$Fe/H$]$ = $-2.25$, $-1.35$, $-0.33$, 0.0 and 0.35 (dashed lines). The
  $V$ and $I$ data are from VLT/FORS1 imaging, while $K$-band data are from 
  NTT/SOFI. Filled circles mark clusters with spectroscopy, colour coded
  as in Fig.~\ref{fig:cmd_vi}. The arrow indicates the foreground
  reddening vector.
}
\label{fig:vi_vk}
\end{figure}

\begin{figure}
\includegraphics[width=85mm]{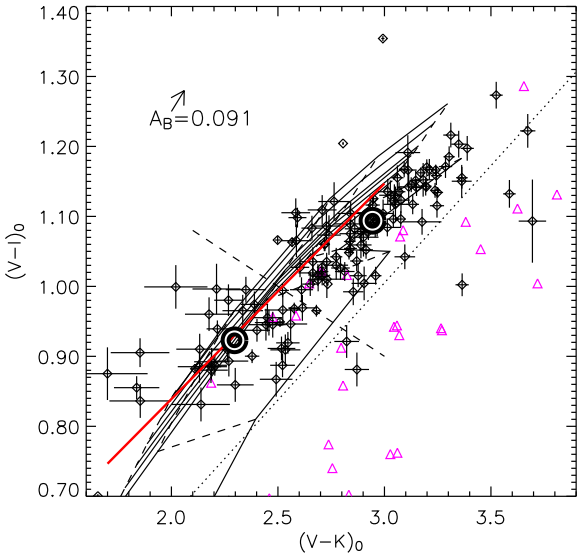}
\caption{As Fig.~\ref{fig:vi_vk}, but with extended and compact sources
  shown with different symbols. Compact sources are shown with error bars
  while extended sources are shown as triangles.
  The two circles mark the average colours of unresolved objects above and
  below the dashed line $(\vi)_0 = -0.2 (\vk)_0 + 1.5$. The solid line
  is the $(\vi)_0$ vs.\ $(\vk)_0$ relation for Galactic GCs, based
  on the colour-metallicity relations in Barmby et al.\ (\cite{bar00}).
}
\label{fig:vi_vk_io}
\end{figure}

\begin{figure*}
\includegraphics[width=170mm]{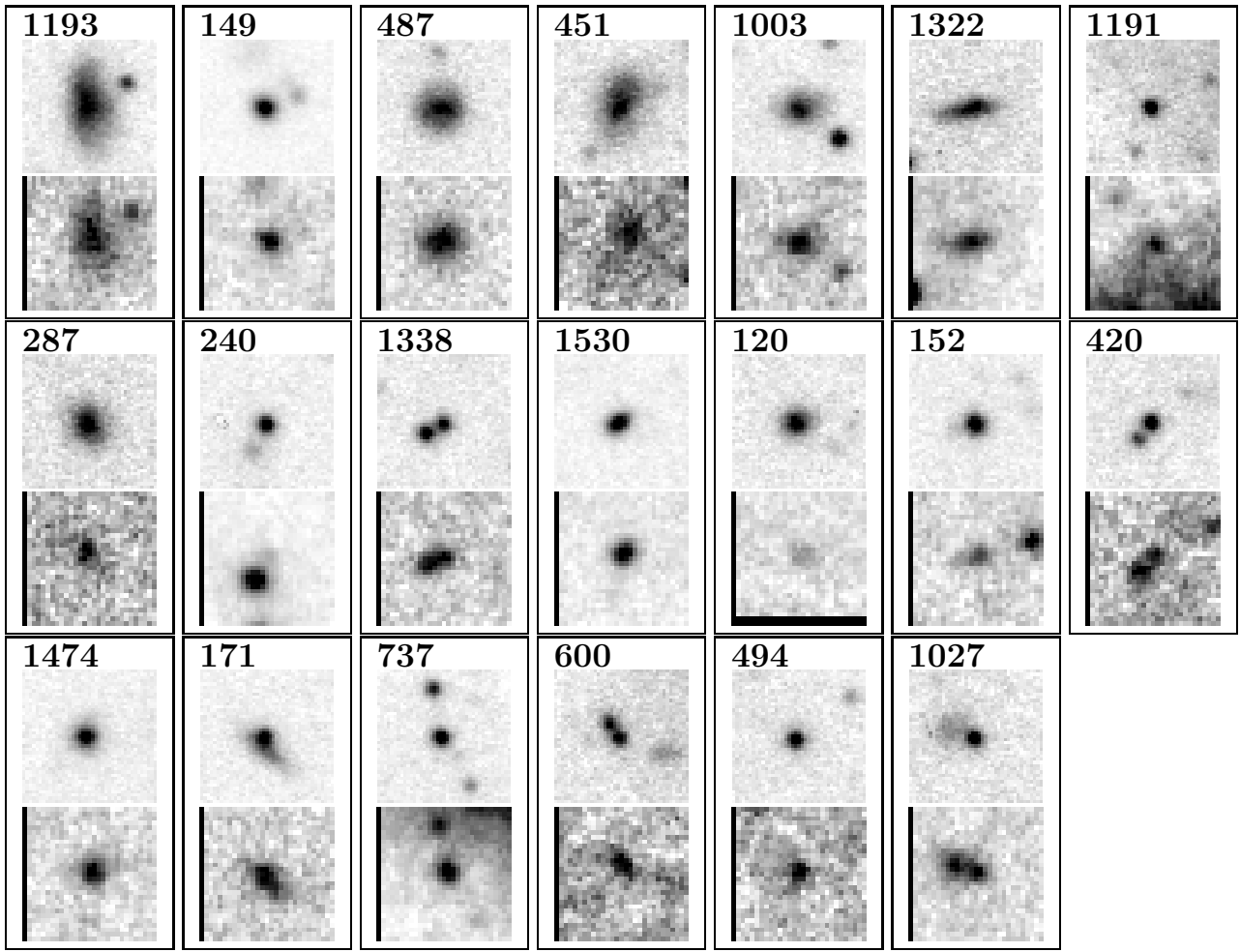}
\caption{Objects with very red \vk\ colours, i.e.\ appearing below the dotted 
  line shown in 
  Fig.~\ref{fig:vi_vk_io}. For each object, the upper and lower panels show the
  FORS1 $V$-band and SOFI $K$-band images.  Starting from the upper left, 
  the objects are arranged in order of decreasing distance from the line.
  Each box subtends $8\arcsec\times8\arcsec$.
}
\label{fig:stamps}
\end{figure*}

  We restrict the following analysis to objects brighter than $V=22.5$,
for which the random photometric errors in all bands are less than about
0.1 mag (Fig.~\ref{fig:perr}).  As a result, all of the GC candidates
are brighter than $K=21$ and about 85\% of the sample fall in the
magnitude range $17<K<20$ (cf.\ Fig.~\ref{fig:kcmp}) with a peak around 
$K\sim19$. 

Fig.~\ref{fig:vi_vk} shows the 
(\vk, \vi) two-colour diagram for the combined SOFI/FORS1 data,
corrected for foreground extinction (Schlegel et al.\ \cite{sch98}).
Also shown are SSP models from Maraston (\cite{mar04}) for ages
between 2 Gyrs and 14 Gyrs and metallicities $-2.25<\feh<0.35$. Like P02
we note that most of the data points fall to the right/below the 
model tracks for old ages, and some data points scatter 
towards the lower right-hand part of the plot. Objects with spectroscopy
from Paper I and II are shown with filled circles. Many of the data points, 
including those representing clusters with spectroscopic data, fall along 
a well-defined, narrow sequence which is about parallel to, but 
offset from the model tracks for old ages and lies between the 2 Gyr and
3 Gyr models.

  To investigate the nature of objects with very red \vk\ colours, 
we drew a line at $(\vi)_0 = 0.335 \, \times \, (\vk)_0$ (shown
as a dotted line in Fig.~\ref{fig:vi_vk_io}). The FORS1 $V$-band and SOFI 
$K$-band images of objects falling below this line are shown in 
Fig.~\ref{fig:stamps} in order of decreasing distance from the line. 
At the distance of NGC~4365, one arcsec corresponds to a linear scale of 
about 100 pc, so most GCs are unresolved at ground-based resolution.
Few if any of the objects in Fig.~\ref{fig:stamps} 
are likely GC candidates - nearly all are clearly 
extended, have close companions, or in a few cases are located close to 
the centre of NGC~4365 (ID 1191 and 737). 

  A visual inspection inspection of the FORS1 images reveals that the 
field is indeed rich in background galaxies, and a further selection of
GC candidates based on size was therefore deemed necessary. To this end
we measured the sizes of each object in the FORS1 $V$-band images
using the ISHAPE code (Larsen \cite{lar99}), assuming King (\cite{king62}) 
profiles with concentration parameter $r_{\rm tidal}/r_{\rm core} = 30$. The 
input PSFs for ISHAPE were generated with DAOPHOT,
using bright unresolved objects in the field.  The ISHAPE size estimates
are listed in the last column of Table~\ref{tab:vik}.

  Fig.~\ref{fig:vi_vk_io} shows the (\vk, \vi) diagram after the
size selection.  We rejected objects with an intrinsic FWHM$>$0.5 
pixels or $\sim10$ pc (i.e.\ corrected for the PSF), which corresponds
to a half-light radius of about 15 pc for the assumed King profiles.
Objects which passed the size criterion are shown with diamond symbols
and error bars while the rejected objects are shown with triangles. 
The larger errors (especially in \vk) for the bluer clusters
are due to the fact that these tend to be fainter overall (cf.\ 
Fig.~\ref{fig:gzcmd} and \S\ref{sec:peccol}), and in $K$ in particular
(as a consequence of their blue colours).
There is a strong preference for the rejected objects to fall below the dotted 
line, leading us to the conclusion (reinforced by Fig.~\ref{fig:stamps}) that 
most of the objects with very red \vk\ colours for a given \vi\
colour are likely background interlopers, as also noted by P02.

  The circles in Fig.~\ref{fig:vi_vk_io} show the mean
colours of ``blue'' and ``red'' clusters, where the dividing line
(shown as a dashed line) is
at $(\vi)_0 = -0.2 (\vk)_0 + 1.5$.  
Additional GC selection criteria
are $0.80 < (\vi)_0 < 1.30$, bluer \vk\ colours than the dotted line, 
and $1.6< (\vk)_0 < 3.8$.
The mean $VIK$ colours of the blue and red GCs are 
$\langle(\vi)_0\rangle_{\rm blue}=0.922\pm0.007$, 
$\langle(\vk)_0\rangle_{\rm blue}=2.296\pm0.037$ and
$\langle(\vi)_0\rangle_{\rm red}=1.094\pm0.006$, 
$\langle(\vk)_0\rangle_{\rm red}=2.944\pm0.023$.  
The errors do not include uncertainties on the photometric zero-points.
We have \emph{not} applied individual 
weights when computing the average colours. This would cause red objects 
(with better S/N in $K$) to be weighted systematically more strongly, 
resulting in a biased estimate of the mean.  

Both the red and blue mean colours are shifted towards redder \vk\ and/or 
bluer \vi\ colours with respect to the oldest isochrones, but the offset
is larger for the red clusters.  
There may be a small group of about 10 objects with intermediate colors 
around $(\vk)_0, (\vi)_0 \approx (2.7, 1.1)$ which are somewhat offset 
from the main group and do appear quite consistent with old ages. An
inspection of the images does not reveal anything special about these
objects other than that all except one are located within $100\arcsec$
of the nucleus of NGC~4365.  

\subsection{Comparison with NGC~3115}
\label{sec:n3115cmp}

\begin{figure}
\includegraphics[width=85mm]{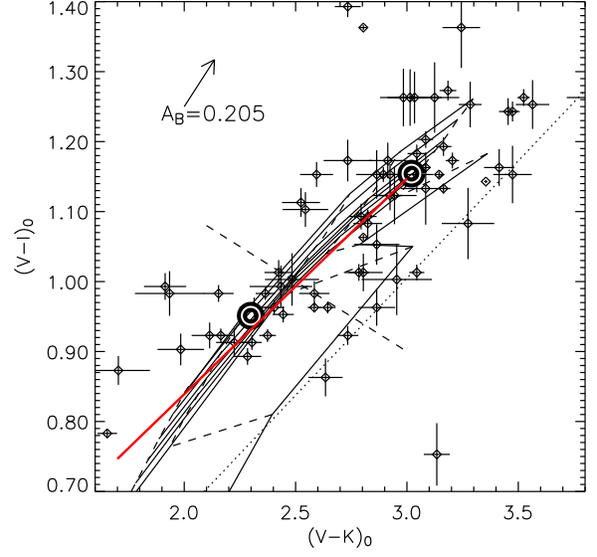}
\caption{As Fig.~\ref{fig:vi_vk}, but for GCs in NGC~3115. The data are
  from P02.
}
\label{fig:vi_vk_n3115}
\end{figure}

\begin{figure}
\includegraphics[width=85mm]{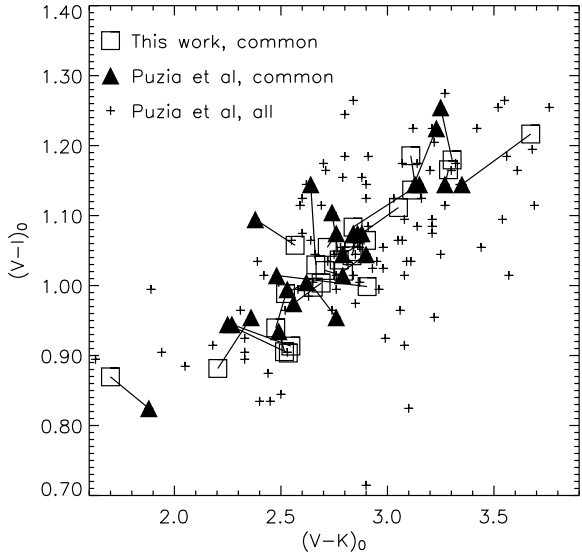}
\caption{Comparison of our $VIK$ data for GCs in the central region
  of NGC~4365 with those of P02.  The 
  open squares and filled triangles mark our photometry and that of
  P02 for objects in common between the two samples. The
  plus markers indicate all objects in the P02 sample. There
  is generally good agreement between the two datasets, but the
  P02 datapoints which are not included in the SOFI/FORS1 data
  (being fainter and/or closer to the centre) show a larger scatter.
}
\label{fig:puz_sl}
\end{figure}

The GC system of NGC~3115 was studied spectroscopically by Kuntschner et 
al.\ (\cite{kunt02}) who found the two GC populations there to be old and 
coeval within 2--3 Gyr uncertainties.  Fig.~\ref{fig:vi_vk_n3115} shows 
the (\vk, \vi) plot for GCs in NGC~3115, using data from P02.  
A correction for foreground extinction
of $A_B=0.205$ mag (Schlegel et al.\ \cite{sch98}) has been applied and
we have used the same size cuts as P02, i.e.\ objects with
FWHM$>0\farcs4$ measured on the HST images were rejected. As in 
Fig.~\ref{fig:vi_vk_io} we show the mean colours of blue and red clusters
with circles, using the same dividing line and selection criteria.
For NGC~3115 these mean colours are
$\langle(\vi)_0\rangle_{\rm blue}=0.951\pm0.009$, 
$\langle(\vk)_0\rangle_{\rm blue}=2.299\pm0.056$ and
$\langle(\vi)_0\rangle_{\rm red}=1.154\pm0.013$, 
$\langle(\vk)_0\rangle_{\rm red}=3.020\pm0.041$.
Although the formal errors on these mean values are
relatively small, there is a large scatter, both real and due to random 
errors, and the representation of the red GCs as a single population may
be an oversimplification, especially in NGC~4365 (see below). 

Comparing Figs.~\ref{fig:vi_vk_io} and \ref{fig:vi_vk_n3115}, the overall
difference between the $VIK$ two-colour diagrams for NGC~3115 and NGC~4365 
does not appear as striking as in the study by P02 (their Fig.\ 6). The 
main difference is that the region corresponding to the most metal-rich 
($Z\ga Z_{\odot}$), old (age$>$5 Gyrs) model tracks is not populated at all 
in NGC~4365, while NGC~3115 has some clusters in this part of the two-colour 
diagram. However, many of these have large error bars and there are also 
objects which scatter to quite young (according to the models) ages.  
The mean colours of the blue GCs in NGC~3115 coincide nicely with
the SSP models for old ages, but the red GCs show an offset (though
smaller than in NGC~4365) with
respect to the models also in this galaxy.  Generally, one gets the 
impression that the NGC~3115 two-colour diagram has more scatter, which 
makes it difficult to compare the two plots directly. This is in spite of 
the combination of HST imaging and ISAAC on the VLT for NGC~3115, compared 
to our FORS1/SOFI data (the exposure times in $K$ are nearly identical). 

The offset between GC colours and models for old ages is larger
in NGC~4365 than in NGC~3115, but this appears to be due to a systematic 
shift affecting both the blue and red GCs.  Interpolating in the 
13-Gyr isochrones, the offsets in \vk\ between models and the average GC 
colours (for fixed \vi\ colour) are 
$\Delta(\vk)_{\rm blue} = 0.07\pm0.04$ mag 
and
$\Delta(\vk)_{\rm red} = 0.28\pm0.02$ mag for NGC~4365. For NGC~3115 we find
$\Delta(\vk)_{\rm blue} = 0.00\pm0.06$ mag 
and
$\Delta(\vk)_{\rm red} = 0.15\pm0.04$ mag.
The difference 
$\Delta(\vk)_{\rm red} - \Delta(\vk)_{\rm blue}$ is about 0.2 mag
in both galaxies, specifically $0.20\pm0.04$ mag in NGC~4365
and $0.15\pm0.07$ mag in NGC~3115.  
A shift of about 0.1 mag 
towards bluer \vk\ colours would make both the red and blue GCs in NGC~4365 
show essentially the same average offset with respect to the 13-Gyr models 
as those 
in NGC~3115.  Such a shift can be easily accommodated when uncertainties on 
the photometric zero-points and the overall scatter are taken into account.
In particular, we remind the reader that our SOFI $K$ magnitudes are
systematically brighter than those based on the ISAAC data of P02
(\S\ref{sec:sofical}), and
an attempt to bring the SOFI magnitudes in closer agreement with the ISAAC
data could shift the \vk\ colours towards the blue by a sufficient
amount to make the red GC populations appear virtually coeval in
the two galaxies. 
 For NGC~3115, the larger scatter means that the average colours depend
more strongly on the exact selection criteria than in NGC~4365. In
particular, there are
more GC candidates falling below the dotted line and having $(\vi)_0>1.3$,
and many objects in the P02 catalog have \vk\ colours outside the plotted
range. If the selection criteria are modified such that no colour cuts 
are applied, the difference $\Delta(\vk)_{\rm red} - \Delta(\vk)_{\rm blue}$
is $0.19\pm0.12$ mag for NGC~3115. For NGC~4365, we get
$\Delta(\vk)_{\rm red} - \Delta(\vk)_{\rm blue} =
0.16\pm0.08$ mag if the colour cuts are omitted.

It is sometimes overlooked that the foreground extinction corrections are 
also uncertain - we have used the Schlegel et al.\ (\cite{sch98}) value of 
$A_B=0.205$ mag for NGC~3115, but the Burstein \& Heiles (\cite{bh82}) value 
is only $A_B=0.100$ mag (both from NED). For NGC~4365 the corresponding values 
are $A_B=0.091$ mag and 0 mag. 
Adopting different foreground extinctions
could shift the mean colours around by a few times 0.01 mag, although
the direction of the reddening vector means that the offsets with
respect to the SSP models will be much less affected.

\subsection{Comparison with Puzia et al.}
  
  In Fig.~\ref{fig:puz_sl} we compare the P02 ISAAC/HST and our SOFI/FORS1 
data for NGC~4365 directly.  The datapoints for the smaller field in 
Fig.~\ref{fig:puz_sl} appear somewhat more concentrated towards intermediate 
colours compared to the full sample in Fig.~\ref{fig:vi_vk_io}. This makes
the well-defined narrow GC sequence in the full sample harder to
recognise, although the objects in Fig.~\ref{fig:puz_sl} do fall along the 
same locus.  This difference suggests a deficiency of the most metal-rich 
and metal-poor objects in the central regions, as discussed further
below (\S\ref{sec:radial}).
The open squares and filled triangles in 
Fig.~\ref{fig:puz_sl} show our photometry and that of P02 for objects in 
common between the two samples. No systematic differences are seen between 
the two sets of measurements, although the scatter may be slightly larger 
for the ISAAC/HST than for the SOFI/FORS1 data. 
Recall that this excellent agreement is a result of similar offsets of 
about 0.16 mag between our $V$ and $K$ magnitudes and those of P02 cancelling 
out when the \vk\ colour is formed (\S\ref{sec:sofical} and 
\S\ref{sec:wfpc2cal}).
The scatter in \vi\ around the 
best-fitting straight line is 0.06 mag and 0.05 mag for the two samples.
The plus markers indicate all GC candidates in the P02 data, 
which show an even larger scatter (0.16 mag).  This is most likely 
because these objects tend to be fainter and/or located closer to the 
centre of the galaxy, where the background is higher. 

\subsection{Intermediate ages? Calibration and Model Uncertainties}

  Before interpreting the above results in terms of age differences,
uncertainties 
in the models need to be carefully considered.  Comparing SSP models from a 
variety of sources, P02 and Hempel \& Kissler-Patig (\cite{hem04a}) found 
substantial differences, but none of the existing models are in 
agreement with the colors of GCs in NGC~4365 if the GCs are indeed old.
Maraston (\cite{mar04}) compares her models with observations of
Milky Way globular clusters and finds generally good agreement, but 
the comparison of \vk\ colours at metallicities $\feh>-1$ is limited by
the small number of suitable objects and a shift of 0.1--0.2 mag
in the \vk\ colour at $\feh\approx0$ does not appear to be ruled out
(her Fig.~21). A shift of about 0.2 mag in \vk\ at the metal-rich end 
would make the (\vk,\vi) 
two-colour diagram consistent with old ages for virtually \emph{all} GCs 
within the SOFI field.

  Would a substantial shift towards redder \vk\ colours make the SSP 
models incompatible with other existing data?
  Hempel \& Kissler-Patig (\cite{hem04a}) show (\vk,\vi) two-colour
diagrams for GCs in 6 galaxies. These include NGC~4365 and NGC~3115 (same 
data as P02), ISAAC/HST observations of NGC~5846 and NGC~7192, and 
Keck/NIRC and WFPC2 data for NGC~4478 and M87.  The M87 and NGC~4478 data 
include only relatively few objects because of the small 
($38\arcsec\times38\arcsec$) NIRC field of view and relatively 
short exposure times of about 10 min for each of two fields on M87 and 
55 min for each of three fields on NGC~4478 
(Kissler-Patig et al.\ \cite{kbm02}).  The largest number of clusters were 
detected in NGC~5846, which shows a similar offset with respect to the 
models as NGC~4365.  The colour distributions in the remaining three
galaxies appear similar to that of NGC~3115, all showing substantial
scatter. The conclusion of Hempel \& Kissler-Patig (\cite{hem04a}) is that 
intermediate-age populations are present in both NGC~4365 and NGC~5846,
while the data for the remaining galaxies are too sparse to put strong
constraints on the age distributions. However, redder \vk\ model colours 
would make the GCs in both NGC~4365 and NGC~5846 appear mostly old, and 
does not appear to be strongly incompatible with the data for the 
remaining galaxies.
  
  Barmby et al.\ (\cite{bar00}) derived relations between various
broad-band colours and spectroscopic metallicities of Milky Way globular
clusters.  They find $\feh = 4.22 (\vi)_0 - 5.39$
and $\feh = 1.30 (\vk)_0 - 4.45$, which can be combined to a relation
between $(\vi)_0$ and $(\vk)_0$: $(\vi)_0 = 0.308 (\vk)_0 + 0.223$.
This relation is drawn as solid straight lines in Figs.~\ref{fig:vi_vk_io} 
and \ref{fig:vi_vk_n3115}. In both NGC~4365 and NGC~3115, the mean
colours of the metal-poor clusters fall very close to this line. In
NGC~3115 the agreement is also excellent for the metal-rich clusters,
while a small offset remains in NGC~4365. However, the Barmby et al.\
colour-metallicity relations themselves are also affected by the
lack of suitable calibrators at high metallicities.

\subsection{Radial trends}
\label{sec:radial}

\begin{figure}
\includegraphics[width=85mm]{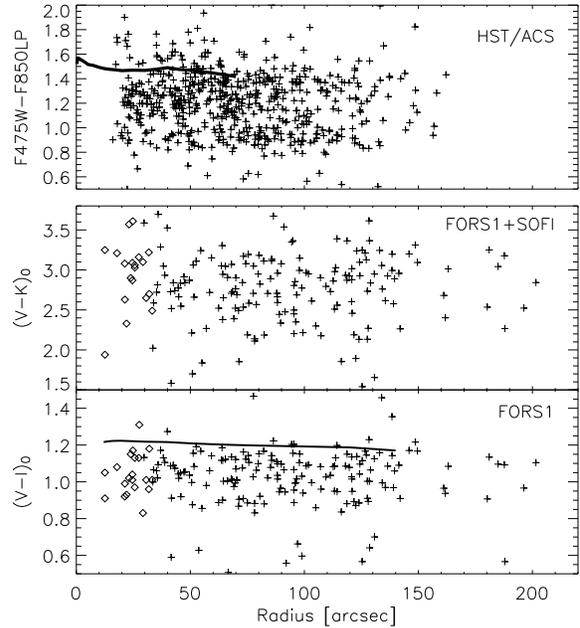}
\caption{\gz, \vk\ and \vi\ colour as a function of projected radial distance 
  from the centre of NGC~4365. The \gz\ colours (top panel) are shown for 
  objects with m(F850LP)$<$23.5. \vk\ and \vi\ colours are for objects with 
  $V<22.5$ and PSF-corrected FWHM$<$0.5 pixels on the
  FORS1 $V$-band image. For radii less than 35$\arcsec$ we are plotting
  $VIK$ data from P02. The solid lines indicate
  the colours of NGC~4365 itself.
}
\label{fig:vi_r}
\end{figure}

\begin{figure}
\includegraphics[width=85mm]{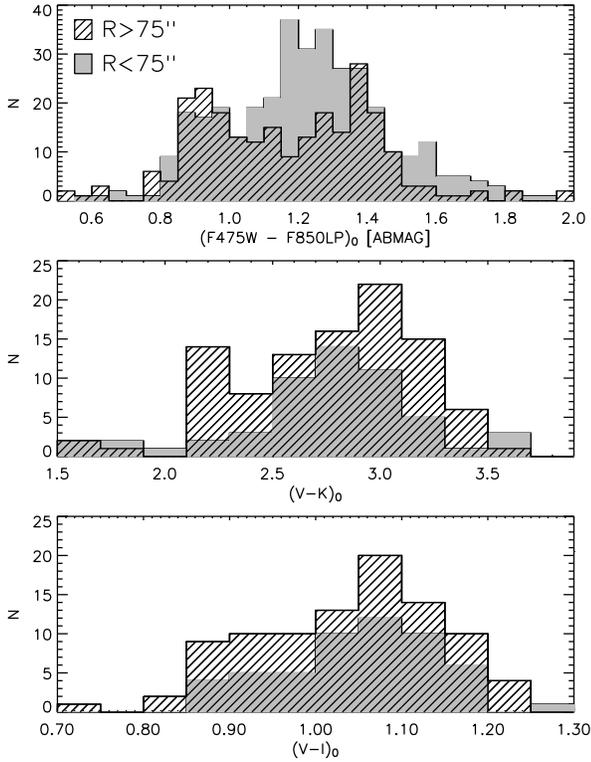}
\caption{Histograms of the \gz, \vk\ and \vi\ colour distributions for
  GC candidates in NGC~4365, divided into ``inner'' and ``outer'' samples.
  Objects inside and outside 75$\arcsec$ are shown with solid filled
  and line-filled histograms. Intermediate-colour GCs are found
  preferentially at small radii.
}
\label{fig:colcmp}
\end{figure}

  In Fig.~\ref{fig:vi_r} we show the \gz, \vi\ and \vk\ colours of GC 
candidates in NGC~4365 as a function of projected radial separation from the 
centre of the galaxy. Selection
criteria are the same as in Figs.~\ref{fig:gzhist} and \ref{fig:vi_vk_io}.
For radii less than $35\arcsec$ we are plotting $VIK$ data from 
P02. Note also that the \gz\ data do not
include the central $\sim20\arcsec$ due to the limit on surface brightness.
The blue GC sequence is again clearly visible, especially in \gz ,
and appears to become less dominant near the centre.

  The integrated \gz\ and \vi\ colours of NGC~4365 itself, measured on the 
ACS and FORS1 images, are also shown (solid lines).
These should only be taken as approximate, since the limited field sizes 
do not allow
us to measure the true sky background. We have used the largest possible
aperture radii for the background measurements 
($r=70\arcsec$ and $r=180\arcsec$ for ACS and FORS1) and the
photometric transformations from \S\ref{sec:fors1}. Within
30$\arcsec$ we get $(\vi)_0 = 1.221$ mag, in virtually perfect agreement with
the value in Table~\ref{tab:gdat}. It is clear that the 
entire GC system remains bluer than the galaxy light at all radii where 
we can measure it.

  In Paper II and in the discussion of Fig.~\ref{fig:puz_sl} we noted a hint 
of a larger concentration of intermediate-colour objects within the central 
parts of the galaxy. This is not easily confirmed by looking at 
Fig.~\ref{fig:vi_r}, but becomes clearer in Fig.~\ref{fig:colcmp} which
shows the
\gz, \vk\ and \vi\ colour distributions for an ``inner'' and ``outer''
sample, divided at $r=75\arcsec$. The overall ratio of outer to inner
clusters is greater for the ISAAC/SOFI data because of the greater area
covered at large radii by these data.  Apart from that, two main differences 
are seen between the inner and outer samples,
most clearly in the \gz\ colours: First, the blue peak is relatively weaker
in the inner sample, as noted above. Second, the red peak is weighted
towards redder colours in the
outer sample. This is consistent with the claim in Paper II of three 
GC populations with mean colours at $\gz\sim$ 0.90, 1.22 and 1.34,
with the intermediate-colour population dominating at small radii
and the red population dominating further out. The difference between
the inner and outer samples is statistically most significant in 
\gz\ colours, with a Kolmogorov-Smirnov test returning a less than
$10^{-3}$ probability that the two samples could have been drawn from 
the same parent distribution. 

It is worth emphasizing that the offset in the (\vk,\vi) two-colour
diagram between the GC colours and the models for old ages is \emph{not}
driven particularly by the intermediate-colour objects which are 
concentrated near the centre. Instead, nearly all clusters except the 
bluest ones show this offset.

\subsection{Comparison of Spectroscopy and $VIK$ colours}
\label{sec:feh_vik}

\begin{table*}
\caption{\label{tab:specphot}
 Clusters with both spectroscopic and photometric data. 
  No reddening corrections have been applied to the photometry in this table.
  Notes: LAR03-2 = BRO05-6, LAR03-5 = BRO05-8, LAR03-11 = BRO05-13.
  The IDs refer to the entries in Table~\ref{tab:vik}.}
\begin{tabular}{lcccccccc} \hline
Name / ID & (x,y) FORS1 & \feh\ (PCA) & $V$ & $\sigma V$ & $I$ & $\sigma I$ & 
     $K$ & $\sigma K$ \\ \hline
BRO05-2 / 328 & 1309.5, 513.1 & $-1.01$ & 21.901 & 0.010 & 20.844 & 0.010 & 19.237 & 0.042 \\ 
BRO05-3 / 402 & 1337.1, 587.6 & $-0.61$ & 21.632 & 0.009 & 20.567 & 0.009 & 18.843 & 0.033 \\ 
BRO05-4 / 389 & 1181.6, 576.1 & $-0.03$ & 21.855 & 0.010 & 20.674 & 0.010 & 18.648 & 0.027 \\ 
BRO05-5 / 408 & 1111.7, 595.3 & $-0.12$ & 21.493 & 0.008 & 20.367 & 0.008 & 18.356 & 0.021 \\ 
BRO05-6 / 494 & 1209.9, 688.0 & $-1.34$ & 22.263 & 0.015 & 21.312 & 0.018 & 19.380 & 0.056 \\ 
BRO05-7 / 505 & 1091.6, 697.5 & $-1.13$ & 21.981 & 0.012 & 20.977 & 0.014 & 19.531 & 0.063 \\ 
BRO05-8 / 677 & 1222.7, 843.2 & $-0.36$ & 21.199 & 0.007 & 20.109 & 0.007 & 18.421 & 0.024 \\ 
BRO05-9 / 645 & 1079.1, 816.8 & $-0.32$ & 21.106 & 0.007 & 20.028 & 0.007 & 18.212 & 0.022 \\ 
BRO05-11 / 766 & 922.1, 906.1 & $-0.43$ & 22.164 & 0.030 & 21.035 & 0.027 & 0.000 & 0.000 \\ 
BRO05-12 / 858 & 835.9, 961.7 & $-0.52$ & 20.777 & 0.007 & 19.687 & 0.007 & 17.996 & 0.020 \\ 
BRO05-13 / 1235 & 1064.4, 1225.5 & $-0.25$ & 22.062 & 0.014 & 20.946 & 0.015 & 19.274 & 0.066 \\ 
BRO05-14 / 1129 & 793.6, 1148.2 & $-0.46$ & 21.434 & 0.009 & 20.352 & 0.009 & 18.460 & 0.031 \\ 
BRO05-15 / 1301 & 792.9, 1276.5 & $-0.21$ & 21.756 & 0.010 & 20.610 & 0.010 & 18.751 & 0.032 \\ 
BRO05-16 / 1402 & 863.6, 1370.6 & $0.18$ & 21.055 & 0.006 & 19.863 & 0.005 & 17.824 & 0.014 \\ 
BRO05-17 / 1452 & 845.1, 1420.4 & $-0.60$ & 21.529 & 0.008 & 20.478 & 0.009 & 18.763 & 0.030 \\ 
BRO05-18 / 1493 & 738.7, 1467.9 & $-0.20$ & 21.412 & 0.007 & 20.291 & 0.007 & 18.402 & 0.021 \\ 
BRO05-19 / 1564 & 644.3, 1569.4 & $-0.49$ & 21.568 & 0.008 & 20.474 & 0.008 & 18.671 & 0.026 \\ 
BRO05-20 / 1609 & 621.3, 1614.3 & $-0.90$ & 21.241 & 0.006 & 20.243 & 0.006 & 18.604 & 0.026 \\ 
BRO05-21 / 1672 & 506.8, 1688.7 & $-0.73$ & 22.495 & 0.016 & 21.450 & 0.019 & 0.000 & 0.000 \\ 
BRO05-22 / 1749 & 631.1, 1794.5 & $-1.01$ & 22.291 & 0.013 & 21.263 & 0.015 & 0.000 & 0.000 \\ 
BRO05-23 / 1840 & 603.6, 1921.7 & $-1.04$ & 22.028 & 0.010 & 21.021 & 0.012 & 0.000 & 0.000 \\ 
LAR03-1 / 448 & 1313.1, 637.5 & $-0.03$ & 21.896 & 0.011 & 20.780 & 0.010 & 18.870 & 0.032 \\ 
LAR03-2 / 494 & 1210.0, 688.2 & $-1.16$ & 22.263 & 0.015 & 21.312 & 0.018 & 19.380 & 0.056 \\ 
LAR03-3 / 536 & 1194.9, 721.9 & $-1.39$ & 22.116 & 0.014 & 21.177 & 0.017 & 19.521 & 0.069 \\ 
LAR03-4 / 678 & 1296.3, 843.0 & $-1.00$ & 21.937 & 0.011 & 20.898 & 0.012 & 19.188 & 0.046 \\ 
LAR03-5 / 677 & 1222.8, 843.3 & $-0.59$ & 21.199 & 0.007 & 20.109 & 0.007 & 18.421 & 0.024 \\ 
LAR03-8 / 1010 & 1123.4, 1061.3 & $-0.63$ & 19.960 & 0.007 & 18.902 & 0.007 & 0.000 & 0.000 \\ 
LAR03-9 / 1182 & 1177.2, 1187.4 & $-0.57$ & 21.428 & 0.008 & 20.344 & 0.008 & 18.527 & 0.029 \\ 
LAR03-11 / 1235 & 1064.8, 1225.1 & $0.00$ & 22.062 & 0.014 & 20.946 & 0.015 & 19.274 & 0.066 \\ 
LAR03-12 / 1211 & 968.2, 1205.0 & $-0.87$ & 21.275 & 0.009 & 20.224 & 0.010 & 18.401 & 0.045 \\ 
LAR03-13 / 1290 & 917.2, 1263.8 & $-0.90$ & 21.509 & 0.010 & 20.443 & 0.011 & 18.579 & 0.029 \\ 
LAR03-14 / 1383 & 977.8, 1355.4 & $0.23$ & 21.538 & 0.008 & 20.386 & 0.008 & 18.433 & 0.027 \\ 
LAR03-15 / 1358 & 848.8, 1334.4 & $-0.38$ & 21.516 & 0.009 & 20.472 & 0.009 & 18.748 & 0.033 \\ 
\hline
\end{tabular}
\end{table*}

\begin{figure}
\includegraphics[width=85mm]{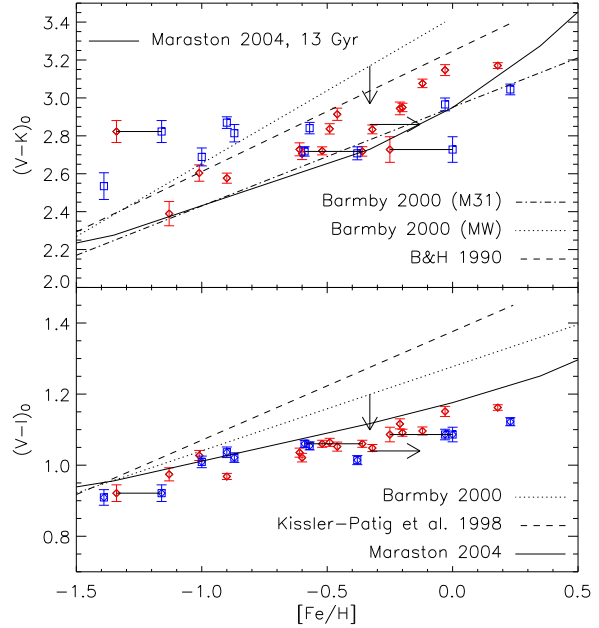}
\caption{\vk\ (top) and \vi\ colour versus spectroscopic metallicity estimates 
 for GCs in the Paper I (blue) and Paper II (red) samples. The solid
 line is a 13-Gyr Maraston (2004) SSP model.  Also shown are
 literature calibrations of \vi\ vs.\ \feh\
 (Kissler-Patig et al.\ \cite{kp98}; 
  Barmby et al.\ \cite{bar00}) and
 \vk\ vs.\ \feh\ (Brodie \& Huchra \cite{bh90} and Barmby et al.\ \cite{bar00}).
 Three objects in common between Paper I and II are connected. The
 arrows indicate the changes in colour and metallicity if
 the clusters are 5 Gyr instead of 13 Gyr old.
}
\label{fig:feh_vik}
\end{figure}

In the following we compare spectroscopic metallicities and broad-band 
colours for GCs in NGC~4365 and two other galaxies, NGC~3115 and the Sombrero
galaxy (\object{NGC~4594}).  For NGC~4365 we use the Paper I and II spectra,
for NGC~3115 we use the Lick index measurements on VLT/FORS2 spectra 
published by Kuntschner et al.\ (\cite{kunt02}) and for the Sombrero we use our
Keck/LRIS data from Larsen et al.\ (\cite{lar02}). Both the Sombrero and
NGC~3115 GC spectra were found by the previous studies to be consistent with 
uniformly old ages for both the metal-poor and metal-rich sub-populations.  
We first look at some general properties of the NGC~4365 data and compare 
these with SSP models and various empirical colour-metallicity relations.
In \S\ref{sec:sdiff} we carry out a differential comparison between the 
different GC systems.

Spectroscopic metallicities were derived using the principal components 
analysis (PCA) method described in Strader \& Brodie (\cite{sb04}, hereafter 
SB04).  SB04 found that the strongest principal component for a set of 
Lick index measurements is strongly correlated with metallicity, and they
established a metallicity calibration based on 39 Milky Way GCs with 
Lick/IDS index measurements from Schiavon et al.\ (\cite{rps04}) and 
metallicities from the McMaster catalogue (Harris \cite{har96}).  According 
to the notes to the catalogue, these metallicities are mostly on the scale 
of Zinn \& West (\cite{zw84}), so the PCA metallicities should also be
roughly on this scale. Table~\ref{tab:specphot} lists
the clusters in NGC~4365 which have both photometric and spectroscopic data.

  In Fig.~\ref{fig:feh_vik} we plot the \vi\ and \vk\ colours versus the 
PCA metallicities for the NGC~4365 GCs. The 13-Gyr Maraston (\cite{mar04}) 
SSP models are drawn as solid curves.  In the upper panel 
we also show the Barmby et al.\ (\cite{bar00}) and Brodie \& Huchra 
(\cite{bh90}) calibrations for \vk\  versus $[$Fe/H$]$, and in the 
bottom panel we show the Kissler-Patig et al.\ (\cite{kp98}) and 
Barmby et al.\ (\cite{bar00}) relations for \vi\ vs.\ $[$Fe/H$]$. The Barmby 
et al.\ relations are based on photometry and spectroscopy of Milky 
Way GCs from Harris (\cite{har96}), so they should also be on the
Zinn \& West scale.  The Kissler-Patig et al.\ relation is based
on Keck spectroscopy of GCs in the giant Fornax elliptical NGC~1399, 
with spectroscopic metallicities derived from the
Brodie \& Huchra calibration which, in turn, is also on the Zinn \& West
scale.  Thus, all the spectroscopic \emph{and}
photometric calibrations can be traced back to the Zinn \& West scale and 
should produce comparable results (whether or not the Zinn \& West 
scale actually measures true Fe abundance).

  While the PCA metallicities and GC colours show a fairly tight 
correlation, the metallicities and the observed \vi\ colours do not agree 
well with any of the empirical colour-metallicity relations in a quantitative 
sense.  The Barmby et al.\ and Kissler-Patig et al.\ 
relations are too red compared to both the Maraston models and the NGC~4365 
data except at the lowest end of the metallicity range.  The 13-Gyr SSP 
models agree better with the GC \vi\ colours, though 
the models remain slightly redder than the GCs or, equivalently, the PCA 
metallicities are too high. This small $\sim0.05$ mag difference might be
partly due to a combination of uncertainties on the photometric
zero-points and the foreground reddening correction, but systematic effects
in the PCA metallicities are also possible (e.g.\ due to resolution effects).

  For \vk\ the empirical relations are also redder than
the metal-rich clusters, but here the Maraston models are actually \emph{bluer} 
than the GCs.  For the metal-poor clusters there is a larger scatter, though 
much of this is driven by one object LAR03-2=BRO05-6 with an anomalously red 
\vk\ colour.  Inspection of the images does not reveal any particular 
complications for this object (nearby neighbours, high background etc.) except 
that it is the faintest object in $V$ and $I$ and the second-faintest in $K$
(Table~\ref{tab:specphot}). 
Part of the difference between data and models could again be due to 
uncertainties on the 
photometric zero-points, in particular on the $K$-band photometry. A shift 
of 0.1 mag towards bluer \vk\ colours would bring the NGC~4365 data closer 
to the models, though still not in perfect agreement. However, errors on
the metallicity calibration cannot simultaneously explain the offsets in
\vi\ and \vk .
Overall, the NGC~4365 GCs appear somewhat bluer in \vi\ and redder in \vk\ 
for their PCA metallicities compared to the Maraston models, consistent with 
the behaviour seen in the (\vk,\vi) two-colour diagrams. 

  The accurate calibration of colour-metallicity relations is 
severely hampered by the paucity of high-metallicity GCs in the Milky Way 
which at the same time have low reddenings. The relations in Barmby et 
al.\ (\cite{bar00}) are based on only a handful of Milky Way GCs with 
$\feh > -1.0$. Including \object{M31} GCs with spectroscopic metallicities 
(based on the calibration in Brodie \& Huchra \cite{bh90}) Barmby et al.\ find 
a shallower \vk\ vs.\ \feh\ relation, indicated by the dotted-dashed line 
in Fig.~\ref{fig:feh_vik}. This relation is in fact \emph{too shallow} to 
fit the NGC~4365 data but agrees rather well with the Maraston (2004)
models. A corresponding relation for \vi\ based on the M31 clusters is
not given.  The substantial differences between the various
colour-metallicity relations mostly likely reflect the lack of suitable 
calibrators at high metallicities.

  Because the spectroscopic metallicity calibrations can be traced back to 
the Zinn \& West (\cite{zw84}) scale for Milky Way GCs, it is implicitly
assumed that the ages of the objects to which these calibrations are
applied are similar to those of GCs in the Milky Way. It is therefore
worthwhile to ask how Fig.~\ref{fig:feh_vik} will change if
the NGC~4365 GCs are intermediate-age. For younger ages,
metallicity sensitive spectral features will generally be weaker and a 
calibration scaled to Milky Way GCs will therefore tend to underestimate the 
metallicities for intermediate-age and young populations. Using the TMB03
models, we estimate that the change in \feh\ (vs.\ only Mg2 and
Fe5270, for simplicity) is nearly linear in age over the range 3--12 Gyr and 
can be approximated as $\Delta \feh / \Delta t({\rm Gyr}) \sim -0.03$
for a metallicity of $\feh=-0.3$ (at 12 Gyr).  Metallicities for a moderately
metal-rich population based on these line indices will then be underestimated 
by about 0.2 dex if applied to a 5 Gyr population.  The same models also
predict that the \vi\ and \vk\ colours for an object with $\feh=-0.3$ will 
change by $-0.1$ mag and $-0.2$ mag between 12 Gyr and 5 Gyr. 

  If the metal-rich GCs in NGC~4365 are 5 Gyr old then the 
metallicities are underestimated and the datapoints in Fig.~\ref{fig:feh_vik} 
should be shifted towards the right by about 0.2 dex. The \vi\ vs.\ \feh\ 
and \vk\ vs.\ \feh\ relations should be shifted down by about 0.1 mag 
and 0.2 mag at the metal-rich end, respectively. These shifts are indicated 
by arrows in the figure.  The net effect of younger ages is modest and 
in the case of \vk\ actually works in the ``wrong'' direction, shifting the 
datapoints slightly further away from the SSP models.
This remains true for all ages greater than 3 Gyr.
For ages between 1 Gyr and 3 Gyr the current models turn redder in \vk\ due
to the TP-AGB phase transition (Maraston \cite{mar04}), but there are
no reliable predictions for the corresponding behaviour of Lick indices.
However, the predicted colour changes are very rapid and even if there
exists an age where the combination of colours and line index strengths
would fit Fig.~\ref{fig:feh_vik} it would presumably require all GCs to be
confined to a narrow age range. This seems unlikely, given the large
metallicity spread, and consequently it appears difficult to explain the 
differences between observed and model colour-metallicity relations as 
a result of young ages.

\subsection{Comparison with Spectra for GCs in Other Galaxies}

  Next, we 
compare \vi\ colours of GCs in the Sombrero galaxy and NGC~3115 with the 
spectroscopic PCA metallicity estimates. For the Sombrero we use the
HST photometry from Larsen et al.\ (\cite{lar02}). For NGC~3115, 
Kuntschner et al.\ (\cite{kunt02}) gave HST/WFPC2 photometry but only
for a subset of the clusters in their spectroscopic sample. Here we
use $V$ and $I$ data from the DFOSC instrument on the Danish 1.54 m
telescope at ESO, La Silla, which were reduced and calibrated using
similar techniques to those described for the FORS1 data in 
\S\ref{sec:fors1}. The data were obtained in December 1999, and
were exposed for about 2 hrs in each of the $V$ and $I$ bands.  This 
provides $V$ and $I$ photometry for 17 GCs in NGC~3115 with spectroscopic
S/N$>$10.

\label{sec:sdiff}

\begin{figure}
\includegraphics[width=85mm]{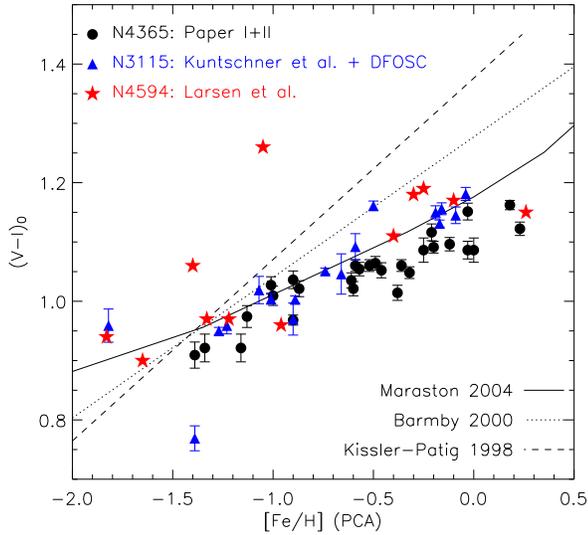}
\caption{\vi\ colour versus spectroscopic metallicity estimates 
 for GCs in NGC~4365 (circles), NGC~3115 (triangles) and NGC~4594
 (stars).  Also shown are the
 Kissler-Patig et al.\ (\cite{kp98}) and Barmby et al.\ (\cite{bar00})
 \vi\ vs.\ \feh\ relations, as well as the 13-Gyr SSP model by
 Maraston (\cite{mar04}). Metallicities were derived using the
 PCA method of Strader \& Brodie (\cite{sb04}).
}
\label{fig:feh_vi}
\end{figure}

Fig.~\ref{fig:feh_vi} shows the \vi\ vs. \feh\ plot for the NGC~4365,
NGC~3115 and Sombrero data.  
All metallicities were derived using the PCA technique. We note at most a 
small overall offset of a few times 0.01 mag between the NGC~4365 and 
NGC~3115 data, although there may be a slight systematic trend in the sense 
that this offset becomes larger at the metal-rich end ($\feh>-0.5$).  The
most metal-rich clusters in NGC~3115 tend to be about 0.05 mag redder
on average than those in NGC~4365 or, equivalently, about 0.3 dex more
metal-poor.  These differences might be ascribed to uncertainties on 
the photometric calibrations and the foreground reddening correction, although
the same trends are seen if we plot the smaller sample of clusters with 
HST/WFPC2 colours from Kuntschner et al.\ (not shown). 
  Note, however, that the FORS1 \vi\ colours for NGC~4365 are slightly redder 
than our WFPC2 photometry, again with a possible slight colour dependence 
(Fig.~\ref{fig:vivi}). If a shift of 0.02-0.03 mag
is applied to the NGC~4365 data then essentially no significant offset is left
with respect to NGC~3115.
Alternatively, there might be small differences in the metallicity scales,
e.g.\ due to the use of different instruments although attempts have been
made to calibrate all measurements to the Lick/IDS system.
The Sombrero clusters show more scatter, which may be partly 
due to reddening internally in the Sombrero. In particular, the outlying 
object at (\feh, \vi) $\approx$ ($-1.05, 1.26$) is located close to a dust 
lane.  

While differences between the datasets in Fig.~\ref{fig:feh_vik} are
marginal, the data for all three GC systems are
clearly offset with respect to
the Barmby et al.\ and Kissler-Patig et al.\ colour-metallicity relations.
In this context, it is interesting to note that Kuntschner et al.\
found their metallicity estimates for NGC~3115 based on $[$MgFe$]$ index 
measurements 
to \emph{agree well} with the \vi\ colours (they compared with the
Kundu \& Whitmore \cite{kw98} calibration for \vi\ vs.\ \feh, which is very 
similar to that of Barmby et al.). Likewise, in Larsen et al.\
(\cite{lar02}) we found the Sombrero GC metallicity estimates based on
the Brodie \& Huchra (\cite{bh90}) calibration to agree well with the
HST/WFPC2 \vi\ colours and the Kissler-Patig et al.\ (\cite{kp98})
relation. Indeed, a comparison of the PCA and Brodie \& Huchra
metallicity estimates for the Sombrero GCs (from Larsen et al.\ \cite{lar02})
shows that the PCA analysis yields systematically higher metallicities at the 
metal-rich end. We find that the relation can be approximated as 
$\feh_{\rm BH} = 0.637 \feh_{\rm PCA} - 0.512$.  If the PCA metallicities
are converted to the B\&H scale, the agreement with the Barmby et al.\
colour-metallicity relation improves substantially with essentially no
offset for NGC~3115 and the Sombrero, while the metal-rich NGC~4365 GCs remain
slightly too blue. A similar transformation would also bring the NGC~4365 data
in better agreement with the Barmby et al.\ \vk\ vs.\ \feh\ relation in the 
upper panel of Fig.~\ref{fig:feh_vik}. Such a correction, however, would make
all the clusters too \emph{red} in both \vi\ and \vk\ compared to the 
Maraston SSP models.

Here we do not intend to imply a preference for one metallicity scale over 
the other, but simply wish to point out that the uncertainties in calibrating 
these relations are still substantial. In a strictly differential sense, 
the NGC~4365 GCs do not appear to behave very differently from their 
counterparts in NGC~3115 and the Sombrero. The possible small offsets towards 
bluer \vi\ colours for NGC~4365 GCs are in the direction expected for younger 
ages, but might also be attributed to calibration differences in colour 
and/or metallicity between the various datasets. Indeed, 
Fig.~\ref{fig:feh_vik} suggests that some systematic differences are present 
even between the two Keck/LRIS datasets for NGC~4365, particularly at 
high metallicities.

\subsection{Spectroscopic ages revisited}
\label{sec:spec_ages}

\begin{figure}
\includegraphics[width=90mm]{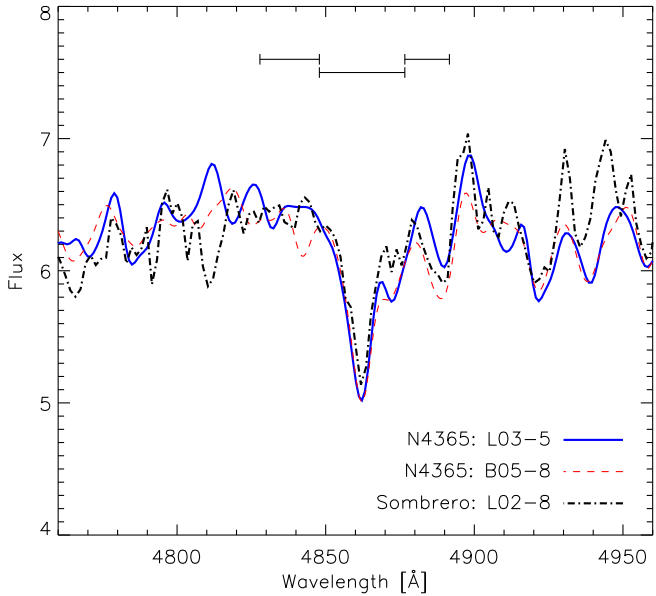}
\caption{Comparison of the Paper I and Paper II Keck/LRIS spectra of 
one GC in NGC~4365 and another GC in the Sombrero of similar metallicity.
The spectra have been smoothed to similar resolutions. The bandpasses
of the Lick/IDS H$\beta$ index are indicated. The slightly deeper
continuum bandpasses in the B05-8 spectrum compared to L03-5 are
responsible for a decrease in H$\beta$ equivalent width from 2.7\AA\ to
1.9\AA.
}
\label{fig:spcmp}
\end{figure}

  Different conclusions regarding the GC
ages were reached based on the two Keck/LRIS datasets. The spectra
presented in Paper I appeared to confirm the presence of intermediate-age
clusters, while the Paper II dataset indicated old ages.  These conclusions 
rest on the ability to measure
Balmer line indices to an accuracy of a few times 0.1 \AA .  In 
Fig.~\ref{fig:spcmp} we show Keck/LRIS spectra from the two datasets for the 
same object, L03-5=B05-8, which showed a large decrease in H$\beta$
equivalent width from Paper I to Paper II. A spectrum for a GC of similar 
metallicity in the Sombrero galaxy is also included.  The figure shows the
region around the H$\beta$ feature for the three spectra smoothed to similar
resolution.  Also shown are the Lick/IDS H$\beta$ feature
and pseudo-continuum bandpasses. For the cluster in NGC~4365 the Paper I 
and II data yield equivalent widths of EW(H$\beta$)=2.7\AA\ and
1.9\AA, while for the Sombrero GC we measure EW(H$\beta$)=1.6 \AA .
The uncertainty from photon noise alone is about 0.3\AA\ in all cases. 

The difference between the Paper I and Paper II H$\beta$ measurements 
corresponds to 
a large shift in the age from $\sim2$ Gyr to $\sim11$ Gyr for this
particular object, but from Fig.~\ref{fig:spcmp} it is clear that the actual 
difference in the spectra is quite small and the stronger H$\beta$ EW in 
the Paper I data actually results from the continuum passbands being slightly 
less depressed rather than any noticeable change in the feature itself. While 
the changes in the line index measurements are uncomfortably large, this is 
one of the few cases where repeated measurements of the same GC have been 
compared. The objects in common between the Paper I and II samples 
were in fact \emph{chosen} to be those having the strongest Balmer line EW
measurements of the Paper I sample, and were thus believed to be the 
best cases for intermediate-age clusters. If instead these strong Balmer
line EWs are simply due to measurement errors, we would also expect these
objects to be the ones showing the largest shifts in a new independent
set of measurements. 

To further investigate the difference between the two datasets, M.\ Beasley
kindly applied the $\chi^2$ fitting analysis by Proctor et al.\ (\cite{pfb04})
to the Paper I and II spectra. This method minimizes the $\chi^2$
difference between all Lick index measurements and Thomas et al.\ 
(\cite{tmk04}) SSP model predictions as a function of age and metallicity.
While the $\chi^2$ method remains subject to possible systematic uncertainties 
in the models, it does provide an independent set of constraints on possible
systematic problems with
spectroscopic data.  For the Paper I data the analysis fails to reach a 
stable solution, and yields ages between 1 Gyr and 15 Gyrs depending on which 
indices are included. For the Paper II data, on the other hand, the analysis
yields consistently old ages ($\sim$15 Gyrs; Paper II). This suggests
that the problem may be with the Paper I spectra, although several
re-reductions of the data have failed to produce any significant
changes in the line index measurements.  However, the Paper I spectra also 
show the largest offset with respect to the 
NGC~3115 data in Fig.~\ref{fig:spcmp}.

\section{Summary and Conclusions}
\label{sec:disc}

  The GC system of NGC~4365 originally gained attention because of its
unusual \vi\ colour distribution.  By now it seems clear that the 
broad-band colour distribution of the GCs is indeed unlike 
those typically seen in other large ellipticals. While the colour distribution
is not unimodal as thought initially, there is a dominant population of 
GCs with intermediate optical colours between the usual ``blue'' and ``red'' 
peaks.
These intermediate-colour clusters may themselves encompass two
sub-populations with slightly different mean colours and the bluer
ones concentrated closer towards the centre of NGC~4365.  The question is 
whether these features are due to a population
of intermediate-age clusters, or clusters which are intermediate-metallicity
but still old.

  Our SOFI data confirm the finding by P02 that the distribution 
of the GCs in the (\vk,\vi) two-colour diagram is shifted towards redder \vk\ 
colours and/or bluer \vi\ colours compared to SSP model tracks for old ages.
If the models are interpreted literally, this makes the majority of the 
intermediate-colour and red GCs within our $5\arcmin\times5\arcmin$ SOFI
field appear as young as 2--3 Gyrs.  
However, we argue that the offset is 
most likely due to a combination of calibration and model uncertainties, 
reducing or eliminating the need to invoke intermediate ages to explain 
the colour distribution.

  As discussed in \S\ref{sec:n3115cmp}, the mean colours of metal-rich GCs in 
NGC~3115 also appear shifted with respect to the models for old ages, 
suggesting that the exact location of the model tracks in the (\vk,\vi) 
two-colour diagram is still rather uncertain.  This is unsurprising, as there 
are few suitable calibrators to test the models empirically. 
If the models are shifted by about 0.2 mag towards redder \vk\ colours at 
the metal-rich end the cluster colours in both NGC~3115 and NGC~4365 would 
be consistent with uniformly old ages. While the metal-rich GCs in NGC~4365 
show a larger offset (by 0.1 mag in \vk) with respect to the models than 
those in NGC~3115, this is also true for the metal-poor clusters.
Accounting also for calibration uncertainties in the $K$-band photometry, 
we are reluctant to take the $VIK$ colours as evidence for an 
intermediate-age population in NGC~4365.

  Our comparison of spectroscopic metallicities and broad-band colours
(\S\ref{sec:feh_vik}) reveals discrepancies between observed GC colours
and models for old ages similar to those seen in the two-colour diagrams.
At a given fixed metallicity, the observed \vk\ colours for GCs in
NGC~4365 are too red (by $\sim$0.2 mag) compared to the Maraston 13-Gyr 
SSP models, while the \vi\ colours are slightly too blue (by $\sim0.05$
mag). Especially the offset in \vk\ colour is very unlikely to be due
to an age difference.

  Compared to the empirical colour-metallicity relations in Barmby et al.\ 
(\cite{bar00}), both the integrated \vi\ and \vk\ GC colours in NGC~4365 are 
too blue at the metal-rich end.  However, the interpretation of these 
differences is unclear as both the spectroscopic and photometric metallicity 
calibrations are uncertain at high ($\sim$ Solar) metallicities.  Bluer 
\vi\ colours compared to spectroscopic metallicity estimates might also be 
expected if the GCs are indeed younger, and we do see hints of a slight offset 
between the colour-metallicity relations in NGC~4365 and NGC~3115.  We doubt 
the significance of this offset, however, as offsets are also seen between 
the two NGC~4365 datasets as well as between the data for GCs in NGC~3115 and 
the Sombrero GCs even though the latter two are both thought to be 
uniformly old.  

  We have compared spectroscopic metallicity estimates with \vi\ photometry 
for GCs in NGC~4365, NGC~3115 and the Sombrero. The colour-metallicity 
relations for the GCs in all three galaxies agree better with the Maraston 
SSP models than with the empirical relations by Barmby et al. However, it
is clear that large uncertainties remain in the calibration of the 
broad-band colour-metallicity relations at high metallicities.

  Considering the uncertainties on the $K$-band zero-points, the 
SOFI/FORS1 \vk\ colours might be too red by up to $\sim0.1$ mag.  A 
correction for this would reduce the need for a shift in the models, but 
probably cannot account for the entire difference.

  The exact origin of the difference between spectroscopic age estimates 
derived in Paper I and II remains unclear. While the Paper II data tend
to have slightly better S/N due to the longer integration times and
better seeing, the difference between the two datasets is clearly 
systematic and not just an effect of larger random errors in either one.
We have performed several re-reductions of the Paper I data and have
been unable to achieve line-index measurements consistent with those
in Paper II. Applying the $\chi^2$ analysis of Proctor et al., we fail
to find a stable solution for the Paper I spectra while the Paper II
spectra are consistent with uniformly old ages.  We note that not only
H$\beta$, but also the higher-order Balmer lines and other features
tend to appear stronger in the Paper I spectra.  Ideally, it would be
desirable to obtain an additional independent set of high S/N spectra
for GCs in NGC~4365.  Currently, we conclude 
that the spectroscopic evidence for an intermediate-age population 
is questionable.

  We briefly summarize the arguments for and against an intermediate-age
GC population in NGC~4365: 

\noindent \emph{Arguments for intermediate-age GCs:}
\begin{itemize} 
  \item (\vk,\vi) and ($U\!-\!I$,\vk) two-colour diagrams show offsets
    between SSP model tracks for old ages and data for NGC~4365, implying 
    ages as young as 2-3 Gyrs if current models are interpreted literally.
  \item Keck/LRIS spectra from Paper I showed Balmer line indices
    suggesting ages in the 2--5 Gyr range (by comparison with
    SSP models) for some clusters.
  \item The red/intermediate-colour GCs extend to brighter magnitudes than
    the blue ones.
  \item The (\vk, \vi) two-colour diagram does show a small number of objects
    which have colours consistent with old ages and 
    intermediate ($\feh\sim-0.33$) metallicities.
\end{itemize} 

\noindent \emph{Arguments against intermediate-age GCs:}
\begin{itemize} 
  \item Strong Balmer lines were not confirmed by Paper II spectra.
        The Proctor et al.\ (\cite{pfb04}) $\chi^2$ analysis, while
	yielding old ages for the Paper II spectra, fails to
	provide robust age determinations when applied to the Paper I data.
  \item SOFI/FORS1 data presented here show that an intermediate-age 
        population, if present, would have to constitute a major fraction 
	of the NGC~4365 GC system. There are very few objects which have 
	colours consistent with an old, metal-rich population.
	This is difficult to reconcile with the 
	uniformly old age for the galaxy itself inferred from spectroscopy 
	and lack of any other evidence for a major event (e.g.\ merger) 
	within the last few Gyrs.
  \item The differences between various empirical and model predictions for
        the GC broad-band colours (especially \vk) as a function of age
	and metallicity are at least comparable
	to the difference between the models for old ages and the
	NGC~4365 data.
  \item A comparison of the offsets between SOFI/FORS1 mean colours of 
        metal-rich and metal-poor GCs in NGC~4365 with respect to the SSP 
	models shows little or no significant difference with respect to 
	data for NGC~3115 whose GCs are believed to be uniformly old. 
  \item Offsets between observed and model colour-metallicity relations
        for the NGC~4365 GCs cannot easily be explained as an age effect but 
	are again consistent with SSP model \vk\ colours being too blue by 
	$\sim0.2$ mag for fixed age at high metallicities. 
\end{itemize}

We suggest that most of the differences between our data for GCs in NGC~4365 
and various theoretical and empirical colour-colour and colour-metallicity 
relations can be attributed to uncertainties in the model- and empirical 
relations, 
combined with general calibration uncertainties on the $K$-band photometry.
We hesitate to interpret the observed properties of the NGC~4365 GC system as 
due to intermediate- or young GC ages.  Better calibrations of spectroscopic 
age and metallicity indicators (e.g.\ Lick system) and integrated colours
(especially those involving near-IR bands) at near-solar metallicities,
as well as comparison with high-quality data for GCs in larger samples
of galaxies, will be required to reach more definitive answers. Systematic 
problems in spectroscopic data also need to be better understood. 

While globular clusters may indeed be useful tracers of the evolutionary 
history of their host galaxies in many cases, NGC~4365 serves as a reminder 
that there may not always be a one-to-one correspondence between GC formation 
and the general stellar population in galaxies. There may be a range 
of field star metallicities in NGC~4365, likely covering about the same 
metallicity range as the GC system and sharing, to a large extent, a similar 
formation history. However, as in the case of NGC~5128, it appears that 
the specific frequency of globular clusters is a strong function of 
metallicity (although that is not necessarily the driving parameter). In 
particular, there can be only a small number of field stars
associated with the intermediate-colour clusters (whether they are young
or not), or otherwise the integrated colours of NGC~4365 itself should
also have been shifted towards the blue. 

Clearly, much more work needs to be done before we can claim to understand 
the relation between GC systems and their host galaxies. Future large
space-based telescopes and 30--100 m ground-based telescopes equipped
with adaptive optics should be able to resolve individual red giant
branch stars in galaxies at the distance of Virgo and obtain
high-dispersion, high S/N spectra of their GCs. Direct comparisons of
field star metallicity distributions with those of GCs will almost certainly
yield several surprises.

\begin{acknowledgements}
We are grateful to Thomas Puzia for sending us his photometry for NGC~3115 
and NGC~4365, to Harald Kuntschner for providing us with a table of his 
line-index measurements in machine-readable form, and to them both for
helpful comments and discussions.  We are also indebted to Mike Beasley 
for applying the Proctor et al.\ $\chi^2$ analysis to the Paper I spectra
and to the La Silla staff for carrying out our SOFI observations in 
service mode. We thank Andr{\'e}s Jord{\'a}n and the anonymous referee for a 
number of useful suggestions which helped improve the paper.
This research has made use of the NASA/IPAC Extragalactic Database (NED) 
which is operated by the Jet Propulsion Laboratory, California Institute of 
Technology, under contract with the National Aeronautics and Space 
Administration. JPB and JS acknowledge support from NSF grant AST-0206139. 
\end{acknowledgements}

\end{document}